\documentclass[aps,twocolumn,longbibliography,superscriptaddress]{revtex4-1}
\usepackage[utf8]{inputenc}
\usepackage{units,stmaryrd}
\usepackage{braket,amssymb,amsmath,graphicx,caption,color,dsfont}
\usepackage[Symbol]{upgreek}
\usepackage{hyperref,xcolor}

\hypersetup{
    colorlinks,
    citecolor=blue,
    filecolor=black,
    linkcolor=black,
    urlcolor=black
}
\newcommand{\diamonds}{\mathbin{\rotatebox[origin=c]{45}{$\square$}}}

\newtheorem{definition}{Definition}
\begin{document}
\title{Superposing compass states
    for asymptotic isotropic sub-Planck phase-space sensitivity}
\author{Atharva Shukla}
\affiliation{%
Institute for Quantum Science and Technology, University of Calgary, Alberta, Canada T2N 1N4
}
\affiliation{%
Department of Physics, Indian Institute of Technology Roorkee,
Uttarakhand-247667, India}
\author{Barry C.\ Sanders}
\affiliation{%
Institute for Quantum Science and Technology, University of Calgary, Alberta, Canada T2N 1N4
}
\date{\today}
\begin{abstract}
Compass states deliver sub-Planck phase-space structure in the sense that sensitivity to phase-space displacement is superior to the sensitivity of displacing the vacuum state in any direction,
but this sensitivity is anisotropic:
better sensitivity for some directions of phase-space displacement vs others.
Here we introduce generalised compass states as superpositions of~$n$ compass states,
with each oriented by~$\nicefrac\pi{2n}$
with respect to its predecessor.
Specifically,
we derive Wigner functions for these generalised compass states
and approximate closed-form expressions for overlaps between generalised compass states and their displaced counterparts.
Furthermore,
we show that generalised compass states,
in the limit~$n\to\infty$,
display isotropic sensitivity to phase-space displacement in any direction.
\end{abstract}
\maketitle

\section{\text{I}ntroduction}
\label{sec:introduction}
Historically,
the coherent state was introduced first by Schr\"{o}dinger~\cite{Schr_dinger1935}
and subsequently,
in seminal work for quantum optics,
by Glauber~\cite{Glauber1963}.
The coherent state,
being a displaced vacuum state,
captures the essence of the position-momentum uncertainty relation ($\Delta x\Delta p\geq\nicefrac\hbar2$) derived from~$[\hat{a},\hat{a}^\dagger]=\mathds1$
for~$\hat a$ the annihilation operator
with $\hat{x}=\hat{a}+\hat{a}^\dagger$
and $\text{i}\hat{p}=\hat{a}-\hat{a}^\dagger$
the position and momentum operators,
respectively.
Inspired by Schr\"{o}dinger's cat-state argument about philosophical implications of quantum mechanics~\cite{Schr_dinger1935},
a quantum optics version of a cat state~\cite{Knight1992},
known at first as even and odd coherent states~\cite{Dodonov1974},
was shown to deliver displacement sensitivity superior to what can be achieved for the coherent state~\cite{Zurek2001,Kumari2015,Toscano2006,Dalvit2006};
this displacement sensitivity technically means that displacing the state in phase space~\cite{Schleich2001},
i.e., applying some combination of position and momentum displacements,
yields a smaller overlap than is achieved by applying the same displacement to a coherent state.

Zurek introduced the compass state,
which is a superposition of four coherent states rotated by~$\nicefrac\pi2$ relative to each predecessor,
which can be pictured as a superposition of coherent states in the north, south, east, and west of phase space relative to the origin.
Then Zurek shows that displacing such a state in any direction of phase space is more sensitive in \emph{every} direction than for the initial coherent state (placed anywhere in phase space)~\cite{Zurek2001},
and this compass state was later connected to 
physical implementations~\cite{Agarwal2003, Pathak2004}
and Heisenberg-limited measurements~\cite{Toscano2006}. Zurek's analysis shows that the sub-Planck phase space sensitivity delivered by the compass state has relevant applications in setting limits on the sensitivity of quantum meters and in accelerated decoherence. In a compass state, the size of the smallest perturbation that makes the displaced state approximately orthogonal and hence distinguishable from the original state is far beyond the `standard quantum limit'. These results also make it possible to anticipate the mesh structure required to simulate a quantum system's evolution in phase space.  

We note that 
this compass state has been generalised to other group symmetries~\cite{Akhtar, Akhtar2022},
but here we generalise in a different way:
we superpose compass states at various orientations and show that, asymptotically for infinitely many such compass states superposed,
such a generalised compass state has sub-Planck phase-space displacement sensitivity that is isotropic with respect to phase-space direction.
This phase-space isotropicity delivers equal sub-Planck resolution regardless of which linear combination of canonical position~$x$ and position~$p$ is being measured. This isotropicity is absent for a single compass state where a perturbation in some phase space direction might make the displaced state approximately orthogonal and hence distinguishable from the original state while not having the same effect when applied in some other phase space direction. 
Our superposition of compass states could be especially useful in the context of quantum metrology.
\par There have been multiple proposals and realisations of sub-Planck phase-space structures. These include the use of dispersive interaction between atoms and a high-quality cavity~\cite{Agarwal2003, Pathak2004}.
Apart from this, there have been multiple actual experimental implementations~\cite{Praxmeyer,Vlastakis,Ofek2016ExtendingTL,Lemos2012,Austin2010}, and their properties have been studied in different contexts~\cite{Jacquod2002,Wisniacki2003,Ghosh2006,PhysRevLett.98.063901,Scott2008,PhysRevA.78.034101,PhysRevA.78.013810,PhysRevA.79.052104,PhysRevA.80.052115,Kumari2015,Dodonov2016,Kumar2017,PhysRevLett.123.020402}. 

In~\S\ref{sec:conceptsnotation} we present salient notions and the notation we use.
We have introduced some compact notation, 
which is quite helpful for us to keeping concepts clear.
Then we present the results in~\S\ref{sec:results}.
We discuss our results in~\S\ref{sec:discussions}
and present our conclusions and outlook in~\S\ref{sec:conclusions}.

\section{Concepts and notation}
\label{sec:conceptsnotation}
Now we discuss the concepts and our notation for cat states and compass states and their generalisation.
We begin with the coherent state and the cat state as a superposition of two coherent states,
and then proceed to discuss and introduce notation for compass states,
which are superpositions of two cat states,
and generalised compass states.
We connect compass states
and generalised compass states,
which are superpositions of compass states,
to the concept of superpositions of coherent states on a circle in phase space~\cite{coherentstatescircles}.
Finally,
we discuss Wigner functions as representations of such states and simplified versions that capture essential features and ignore relatively minor features.

Mathematically,
the coherent state is
\begin{equation}
\label{eq:coherentstate}
    \ket{\alpha}:=D(\alpha)\ket0,\,
    \alpha\in\mathbb{C},
\end{equation}
where 
\begin{equation}
\label{eq:displacement_operator}
    D(\alpha):=\exp\{\alpha\hat{a}^\dagger-\text{hc}\}
\end{equation}
is the Glauber-Sudarshan displacement operator,~$\ket0$ the vacuum state
and hc denoting the hermitian conjugate~\cite{Gazeau2009}.
For simplicity,
we introduce the notation~$\ket a$
for~$a\in\mathbb{R}^+$
as we only require a positive real-valued coherent state for our purposes.
The even and odd cat state is then
\begin{equation}
\label{eq:evenoddcatstate}
\ket{a}_\pm:=\ket{a}\pm\ket{-a},
\end{equation}
and all our states are implicitly normalised so we do not write explicitly the normalisation coefficients.
For convenience we represent the even cat state by~$\ket{\text{\textbf{\textthreequartersemdash}}}:=\ket{a}_+$,
with the amplitude~$a$ suppressed because the amplitude~$a$ is fixed to be the same for all states discussed herein.

Now we show how to extend an even cat state to a compass state and superpositions thereof.
The $\nicefrac\pi2$-rotated version of an even coherent state is expressed simply as
\begin{equation}
\label{eq:evenoddrotcohstate}
\ket{\text{\ \textbar\ }} :=R\left(\nicefrac\pi2\right)\ket{\text{\textbf{\textthreequartersemdash}}},\,
R(\theta):=\text{e}^{\text{i}\theta\hat{a}^\dagger\hat{a}}.
\end{equation}
We denote the compass state by
\begin{equation}
\label{eq:compassstate}
\ket{\diamonds} =\ket{\text{\ \textbar \ }}+\ket{\text{\textbf{\textthreequartersemdash}}},
\end{equation}
which is a special case of coherent states on the circle~\cite{DAJ94,coherentstatescircles}.
Furthermore,
we introduce the convenient rotated-state notation \begin{equation}
\label{eq:rotatedstate}
\ket{\;}_\theta:=R(\theta)\ket{\;}
\implies
\ket\alpha=\ket{a\equiv|\alpha|}_{\arg\alpha}.
\end{equation}
Thus,
\begin{equation}
\label{eq:compassgeneral}
\ket{\diamonds}_{\theta}
:=R(\theta)\ket\diamonds
\end{equation}
with special cases
\begin{equation}
\label{eq:compasscases}
\ket{\diamonds}_{\nicefrac\pi4}:=\ket{\square},\,
\ket{\diamonds}_{\nicefrac\pi6}
:=\ket{{\rotatebox[origin=c]{75}{$\square$}}},\,
\ket{\diamonds}_{\nicefrac\pi3}
:=\ket{{\rotatebox[origin=c]{105}{$\square$}}}.
\end{equation}
A superposition of~$n$ compass states rotated by a fixed amount with respect to the predecessor and equidistributed over phases from~$0$ to~$2\pi$ is
\begin{equation}
\label{eq:ncompassstates}
\sum_{m=0}^{n-1}\ket{\diamonds}_{\nicefrac{m\pi}{2n}},
\end{equation}
which is a special case of superpositions of coherent states on a circle in phase space,
which has been studied in quite general form but not specifically for the restricted case~(\ref{eq:ncompassstates}) here~\cite{coherentstatescircles}.

Wigner functions~$W(x,p)$ are representations of states that help us to visualise quantum states as phase-space representations~\cite{Schleich2001}.
Wigner functions are a close quantum analogue of classical phase-space distributions~\cite{Schleich2001,Navarrete-Benlloch2015}.
For a pure state~$\ket{\psi}$,
\begin{equation}
\label{eq:wignergeneral}
    W(x,p;\psi)
    :=\int_{\mathbb{R}}\text{d}y\,
\text{e}^{\nicefrac{\text{i}py}\hbar}\psi^{*}\left(x+\nicefrac{y}{2}\right)\psi\left(x-\nicefrac{y}{2}\right),
\end{equation}
which is not explicitly normalised per our convention.
As a special case,
the Wigner function for the coherent state~$\ket{a}$~(\ref{eq:coherentstate}) is
\begin{equation}
\label{eq:Wcohstate}
W_\bullet(x,p;a)
=\exp\left(-\left(p^2+\left(x-2a\right)^2\right)/2\right),
\end{equation}
where~$_\bullet$ denotes a coherent state in our short-hand notation.
The Wigner function for the cat state~$\ket{\text{\textbf{\textthreequartersemdash}}}$ is 
\begin{align}
\label{eq:Wcatstate}
W_{\text{\textbf{\textthreequartersemdash}}}(x,p;a)
    =& W_\bullet(x,p;a)+W_\bullet(x,p;-a)\nonumber\\
& + 2\text{e}^{-\left(x^2+p^2)\right/2}\cos{(2ap)},
\end{align}
with the subscript~$_{\text{\textbf{\textthreequartersemdash}}}$ denoting the even cat state.
This cat-state Wigner function~(\ref{eq:Wcatstate})
is the sum of Wigner functions for two coherent states~(\ref{eq:Wcohstate})
along with a third `interference' term.

Now we introduce notation and expressions for the superposition of four coherent states in two cases using the north-south-east-west language explained in~\S\ref{sec:introduction}: the north-south case and the east-west case.

First, we introduce Wigner functions for the Gaussian lobes of the cat states.
Specifically,
the Wigner function
\begin{equation}
\label{eq::}
W_:(x,p;a)
   :=W_\bullet(p,x;a)+W_\bullet(p,x;-a)
\end{equation}
with~$_:$
our short-hand notation to denote the two coherent states at the North and South end of the compass and does not include the interference terms between them.
The next term is
\begin{equation}
\label{eq:..}
W_{\rotatebox[origin=c]{90}{:}}(x,p;a)
   :=W_\bullet(x,p;a)+W_\bullet(x,p;-a)
\end{equation}
with~$_{\rotatebox[origin=c]{90}{:}}$
our notation for two coherent states at the East and West points of the compass and without interference terms.

Now we introduce the remaining terms required to write the Wigner function of a compass state.
First,
we use the notation
\begin{equation}
\label{eq:onecompasscentre}
   W_+(x,p;a)
   :=2\text{e}^{-\left(p^2+x^2\right)/2}\big(\cos\left(2ap\right)+\cos\left(2ax\right)\big)
\end{equation}
for the interference pattern at the centre.
This Wigner function is calculated according to the contribution of the interference terms from the two cat states formed by the diagonally opposite coherent states.
Next, we consider interference terms for adjacent cat states in the northwest, northeast, southwest and southeast directions
and write
\begin{equation}
   W_{\rotatebox[origin=c]{45}{$\boxdot$}}
   :=2\sum_{\sigma_{x},\sigma_{p} = \pm1} G(\sigma_{x}x,\sigma_{p}p)
\end{equation}
with~${\rotatebox[origin=c]{45}{$\boxdot$}}$ being our notation to represent the rhombus and
\begin{align}
\label{eq:G}
G(x,p;a) := &\ \text{e}^{-\nicefrac{1}{2} \left((x-a)^2 + (p-a)^2 \right)} \nonumber \\
& \times \cos\big(a(x+p-a)\big).
\end{align}
Combining all these terms yields the full Wigner function
\begin{equation}
\label{eq:split-up-compass}
    W_{\diamonds} = W_{:}+W_{\rotatebox[origin=c]{90}{:}}
    + W_{\rotatebox[origin=c]{45}{$\boxdot$}}+W_+
\end{equation}
for the entire compass state. The Wigner function for rotated compass states~(\ref{eq:compasscases}) can be found similarly by rotating the coordinate system.
Now we define sensitivity to displacement in phase space and Planck scale.
\begin{definition}[Sensitivity]
\label{def:sensitivity}
For any given small real-valued threshold,
sensitivity of the state is the smallest magnitude of displacement over all phase-space directions such that the overlap of the displaced state with the original state is less than the threshold.
\end{definition}
\noindent
Now we explain phase-space displacement and how it can be used to quantify sensitivity and establish the Planck scale.

To quantify the sensitivity of phase-space displacement on a state, 
we employ the overlap between a state and its $\delta$-displaced version,
namely,
\begin{equation}
\label{eq:overlapdefinition}
\gamma(\delta;\ket{\psi})
:= \left|\bra{\psi}|D(\delta)|\ket{\psi}\right|^2
\Rightarrow\gamma(0;\ket{\psi})=1\forall\ket{\psi}.
\end{equation}
If $\gamma(\delta;\ket{\psi})=0$,
the orthogonality condition
\begin{equation}
\label{eq:perp}
\ket\psi\perp D(\delta)\ket\psi
\end{equation}
holds and thus the two states in Eq.~(\ref{eq:perp})
are perfectly distinguishable.
Sensitivity does not require that the overlap be zero but rather only needs to be small enough with respect to a small real-valued threshold~$\epsilon$.
Thus, sensitivity is quantified by~$|\delta|_\text{min}$,
which is given by~$|\delta|$ satisfying
\begin{equation}
\min_{|\delta|,\text{arg}\delta} \gamma\left(\delta;\ket{\psi}\right)<\epsilon
\label{eq:sensitivity}
\end{equation}
with respect to the given threshold~$\epsilon$.
Thus,
we employ~$|\delta|_\text{min}$
as the sensitivity quantifier in accordance with Def.~\ref{def:sensitivity}. 

We quantify the sensitivity of the state~$\ket\psi$
to displacement by calculating the smallest~$|\delta|$
such that the orthogonality condition~(\ref{eq:perp}) holds.
Smaller~$|\delta|_{\min}$ such that Eq.~(\ref{eq:perp}) holds indicates a more sensitive state.
For the study of the sensitivity of superpositions of compass states,
the expression
\begin{align}
\label{eq:generaloverlap}
    {}_{\theta_1}\!\bra{a}D(\delta)\ket{a'}_{\theta_2}
    =& \exp\left\{
    \text{-i}\big(a\sin{\theta_{1}(a'\cos{\theta_2+\text{Re}(\delta)}})\big)\right\}
    \nonumber\\
    &\times\exp\left\{\text{-i}\big(-a\cos{\theta_1}(a'\sin{\theta_2+\text{Im}(\delta))\big)}\right\} \nonumber \\
    & \times \text{e}^{\text{i}\big(\text{Im}(\delta)a'\cos{\theta_2-\text{Re}(\delta)a'\sin{\theta_2\big)}}} \nonumber \\
    & \times \text{e}^{\nicefrac{-\left|a\text{e}^{\text{i}\theta_{1}}-a'\text{e}^{\text{i}\theta_{2}} - \delta\right|^2}{2}}
\end{align}
is handy later.
The simple version
\begin{equation}
{}_{\theta}\!\bra{a}{D}(\delta)\ket{a}\!{}_{\theta} =\text{e}^{2\text{\text{i}}\big(a\text{Im}(\delta) \cos\theta -a\text{Re}(\delta)\sin\theta\big)-\nicefrac{|\delta|^2}{2}}
\end{equation}
holds for
\begin{equation}
\label{eq:aa'theta}
a=a',\,
\theta_1=\theta_2
\end{equation}
with the squared magnitude being
\begin{equation}
\label{eq:cohstateoverlap}
\left|\bra{a}D(\delta)\ket{a}\right|^2
=\text{e}^{-|\delta|^2}.
\end{equation}
Importantly, the right-hand side of Eq.~(\ref{eq:cohstateoverlap})
is independent of the choice of coherent state.
Now that we have established concepts and notation for compass states, we proceed to study the properties of compass states and its superpositions.
\section{Background on the Compass State}
\label{sec:one compass}
In this section, we review the properties of a compass state~$\ket{\diamonds}$
as knowledge of these properties is vital for our contributions in subsequent sections.
We begin by reporting the Wigner function~(\ref{eq:wignergeneral}) for the compass state
and the overlap of a compass state with its displaced counterpart.
Finally, we report the solution for the phase-space regions where this overlap function is approximately zero.

\begin{figure*}
\includegraphics[scale = 0.25]{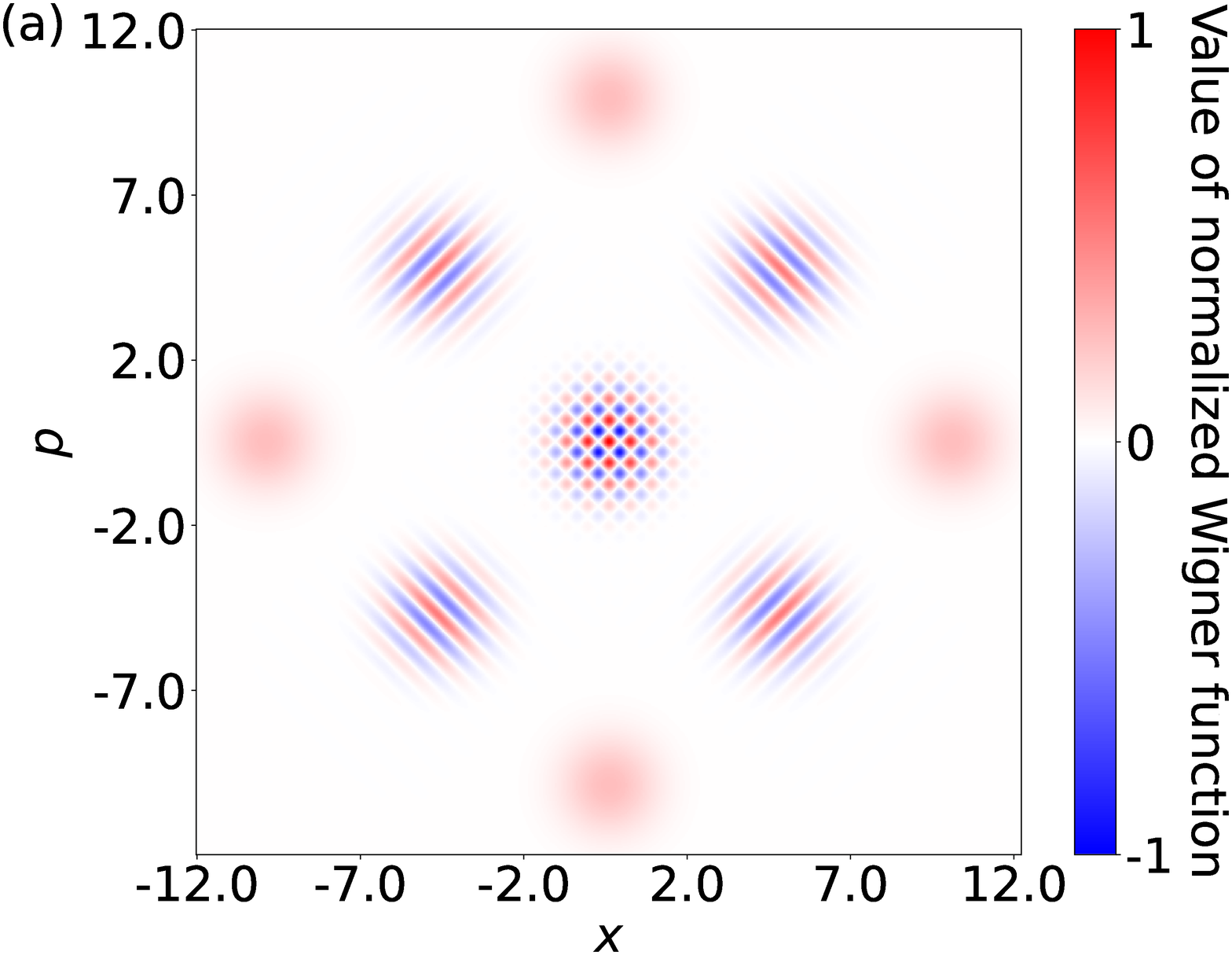}
\includegraphics[scale = 0.25]{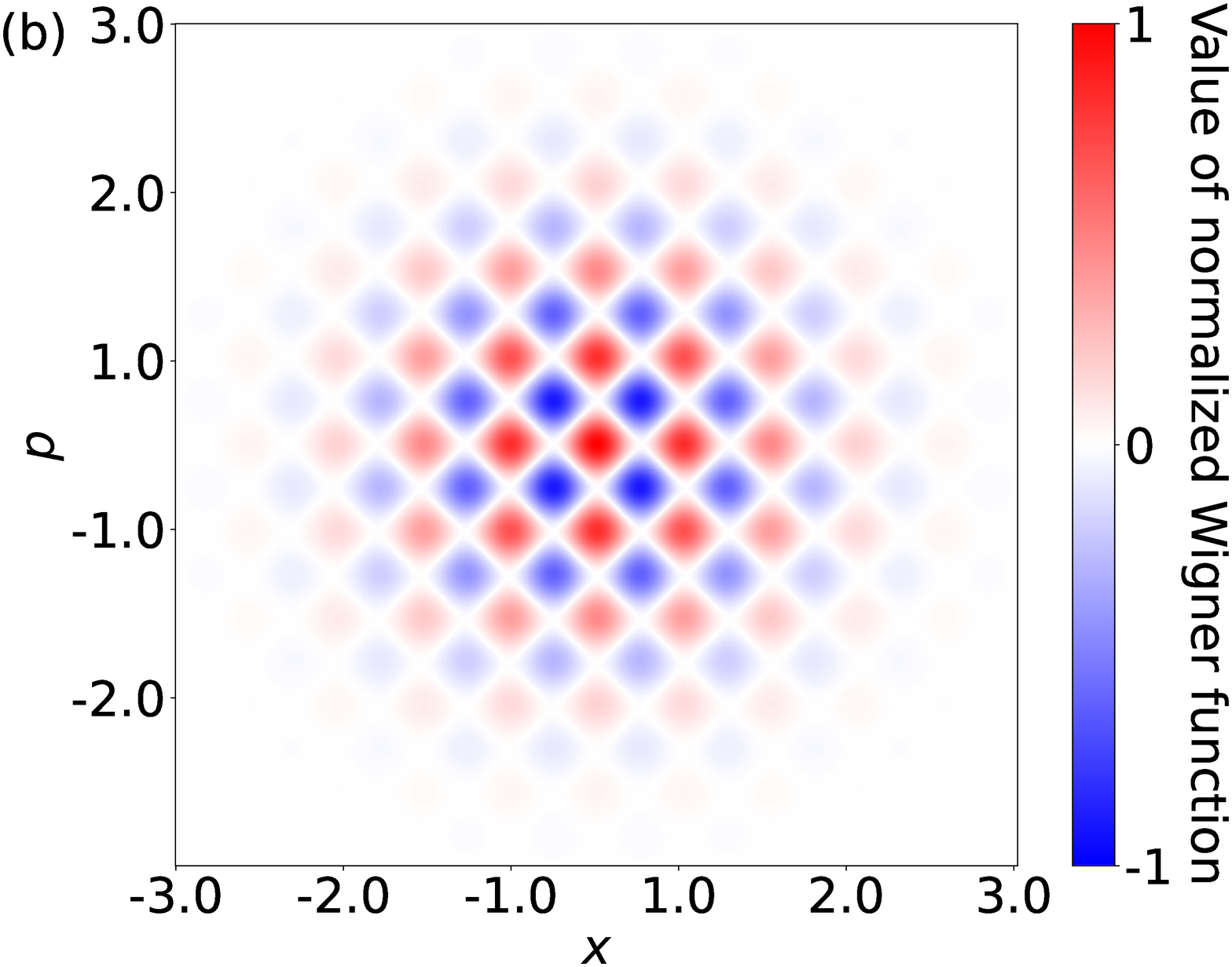}
\caption{%
Heat map for the normalised Wigner function of a compass state with $a=5$ (a)~shown fully and (b)~the centre interference pattern only.
}
\label{fig:WignerOneCompass}
\end{figure*}

\paragraph{}
We present a plot of the Wigner function for the compass state in Fig.~\ref{fig:WignerOneCompass}(a).
The full expression has been discussed in~\S\ref{sec:conceptsnotation}.
We choose $a=5$,
i.e., five times the Planck scale in phase space,
because this choice of~$a$ effectively suppresses the overlap between coherent states (whose size is given by~$a$)
and nearest neighbours.
In this case,
observed phase-space interference arises for distinct coherent states.
Choosing small~$a$ for the compass state would be analogous to forming kitten, rather than cat, states,
corresponding to superpositions of overlapping coherent states~\cite{SUP+17}.
We choose~$a$ such that the distance between the adjacent coherent states of our state is at least 6 standard deviations, i.e. 6 units. 

\paragraph{}
The smallest interference structures arise from coherence between pairs of coherent states maximally separated by angle in phase space, i.e., up to $\pi$, available to them in the phase space~\cite{Zurek2001}.
The linearity of interference means that we do not need to treat interference between all coherent states but rather between pairs,
but note that this approach leads to a factorially growing number of pairwise interference terms so we also have a strategy for reducing an~$n!$ cost to a cost of~$2n$
with~$n$ the number of compass states,
each with~$4$ coherent states superposed.
As we are only concerned with the centre part of phase space,
where sensitivity is the highest,
we focus on the interference term near the centre.
\paragraph{}
Now we explain technically how we calculate the Wigner function for the centre of phase space for a single compass state.
A compass state has four equally spaced coherent states on the circle in
the North (N), South (S), East (E) and West (W) of phase space as introduced in~\S\ref{sec:introduction}.
There are $4\choose2$ cat states in the compass state. The interference pattern near the centre is formed by the contribution of only the pairs of coherent states maximally separated by angle in phase space $(\pi)$, i.e., from the interference of the N-S and E-W pairs of coherent states. We ignore the contributions from the interference of the N-E, E-S, S-W, and W-N cat states. The Wigner-function representation is
\begin{equation}
\label{eq:onecentre}
    W_+(x,p)=2\text{e}^{-\nicefrac12\left(x^2+p^2\right)}\left[\cos\left(2xa\right)+\cos\left(2pa\right)\right].
\end{equation}
This is a sign-alternating pattern; the area of the tile of this chessboard-like pattern is equal to~$\nicefrac{\pi^2}{2a^2}$.
As the area is proportional to~$a^{-2}$, it is smaller than~$\hbar^2$ for~$a\gg1$,
which correspond to sub-Planck structures~\cite{Zurek2001}.
\paragraph{}
A similar alternating chessboard pattern emerges for a mixture of a horizontal and vertical cat state as well. 
However,
sensitivity is enhanced only for certain phase-space directions~\cite{Akhtar},
which casts some doubt over the importance of sub-Planck phase-space structure as a way to assess enhanced sensitivity.
Phase-space displacement sensitivity avoids this problem~\cite{Zurek2001}.

To determine phase-space displacement sensitivity of a compass state,
we need to calculate an approximation of the overlap of a compass state with its displaced counterpart.
The full expression for the compass state is complicated,
and previous work presents the overlap function for a simplified case that ignores cross terms,
corresponding to superpositions of some pairs of coherent states,
with this neglect of cross terms not explicitly stated~\cite{Akhtar}.
Here we calculate the full exact expression~(\ref{eq:exactoverlap}) as we need to consider cross terms carefully in superposing compass states.

\begin{figure*}
\includegraphics[scale = 0.25]{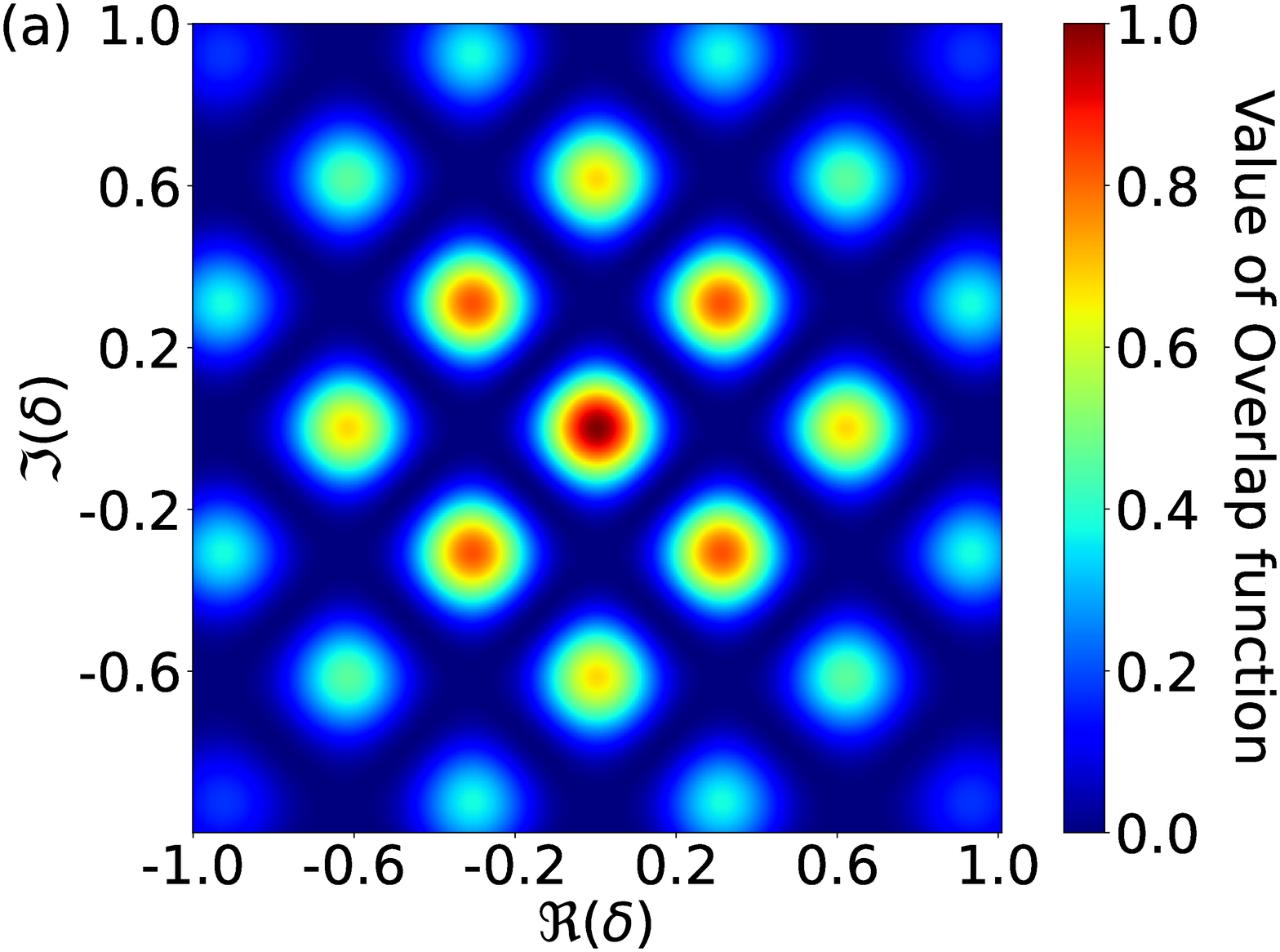}
\includegraphics[scale = 0.25]{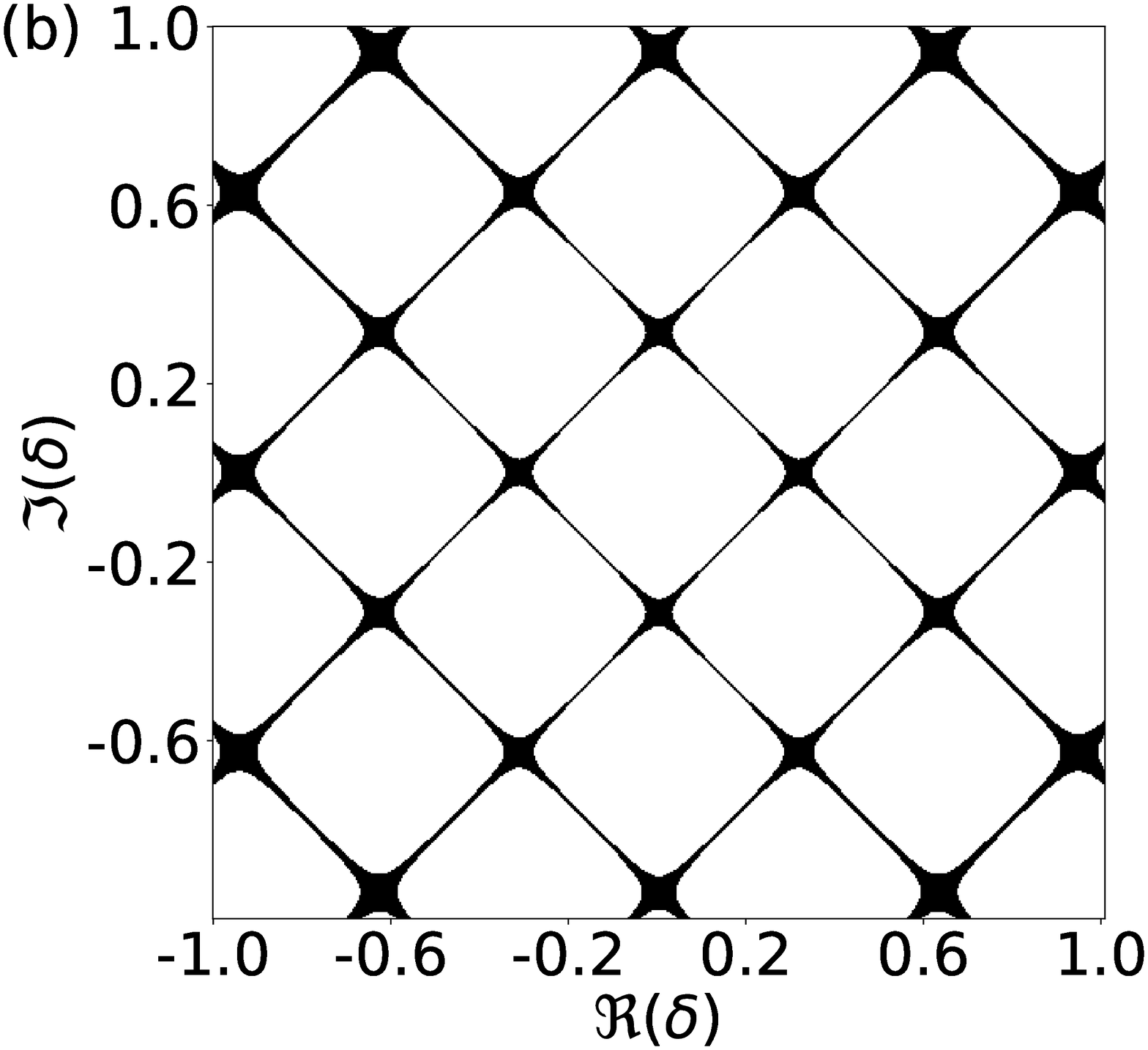}
\caption{Heat map for the overlap function of a compass state with $a=5$ (a)~shown fully and (b)~the phase space regions where its value is approximately zero.}
\label{fig:OverlapFunctionOneCompass}%
\end{figure*}

We now explain the method for ignoring negligible cross-terms using the following terminology:
the original compass state as a superposition of north, south, east and west coherent states in phase space with the displaced compass state comprising a superposition of coherent states at positions north$'$, south$'$, east$'$ and west$'$.
The neglected cross terms correspond to the
\begin{itemize}
    \item north-south$'$ (NS$'$), north-east$'$ (NE$'$), north-west$'$ (NW$'$),
    \item south-north$'$ (SN$'$), south-east$'$ (SE$'$), south-west$'$ (SW$'$),
    \item east-north$'$ (EN$'$), east-west$'$ (EW$'$), east-south$'$ (ES$'$),
    \item west-north$'$ (WN$'$), west-east$'$ (WE$'$) and west-south$'$ (WS$'$)
\end{itemize}
cases,
which contribute negligibly to phase-space displacement sensitivity for sufficiently large~$a$.
As~$a$ increases,
we can see that the cross terms vanish from our derivation of the exact expression~(\ref{eq:exactoverlap}).
The approximate overlap for a compass state with its $\delta$-displaced self is
\begin{equation}
\label{eq:overlaponecompass}
\gamma(\delta;\diamonds)
\approx\text{e}^{-|\delta|^2}\left|\cos\big(2a\text{Im}\left(\delta\right)\big)+\cos\big(2a\text{Re}(\delta)\big)\right|^2
\end{equation}
neglecting the cross terms. 
\paragraph{}
Now we depict the approximate insensitivity of $\gamma$~(\ref{eq:overlaponecompass})
to the direction of displacement~$\delta$ in phase space.
In  Fig.~\ref{fig:OverlapFunctionOneCompass}
we present a plot of $\gamma$~(\ref{eq:overlaponecompass}) vs~$\delta$ and the regions where  $\gamma$ is approximately zero $(<0.001)$.
As seen in Fig.~\ref{fig:OverlapFunctionOneCompass}(b), the region of phase space where~$\gamma$ is close to zero encloses the origin.
The origin signifies the overlap with the unperturbed state, i.e., perfect overlap,
\begin{equation}
\label{eq:perfectoverlap}
\gamma(0;\ket{\diamonds})=1,
\end{equation}
which holds for the exact version of $\gamma$~\eqref{eq:exactoverlap} as well.
Furthermore,
for displacement direction arg$(\delta)$ in phase space,
$\gamma=0$ has infinitely many periodic solutions,
with the period depending on the direction
of displacement.

\paragraph{}
Now we proceed to calculate smallest~$|\delta|$ 
for which~$\gamma$~\eqref{eq:overlaponecompass} vanishes in order to quantify the sensitivity given by Eq.~\eqref{eq:sensitivity} of the compass state,
which corresponds to solving
\begin{align}
\label{eq:overlapvanishtwo}
   \cos\big(2a\text{Im}\left(\delta\right)\big)+\cos\big(2a\text{Re}\left(\delta\right)\big) = 0
\end{align}
whose solutions are
\begin{equation}
    \text{Im}(\delta) \pm \text{Re}(\delta) =\nicefrac{(2m+1)\pi}{2a},\,
    m\in\mathbb{Z}.
\end{equation}

The minimum magnitude of displacement is thus
\begin{equation}
\label{eq:mindisp}
|\delta|_\text{min}=\nicefrac{\pi}{2\sqrt{2}a}, \arg\delta = \nicefrac{(2m+1)\pi}4, m\in\{0,1,2,3\}
\end{equation}
with four solutions corresponding to four coherent states superposed to make a compass state.
Our aim in superposing compass states is to devise a state for which the sensitivity is approximately constant over all $\text{arg}(\delta)\in[0,2\pi)$.
\section{Results}
\label{sec:results}
\paragraph{}
In this section, we study the specific case of the superposition of two compass states in~\S\ref{sec:two compass}. We calculate the Wigner function and solutions for the phase-space regions where $\gamma\approx0$. Subsequently, we generalise our results for an arbitrary number $n$ of compass states in~\S\ref{section:generalised_superposition}.
\subsection{Superposition of two compass states}
\label{sec:two compass}
\paragraph{}
Now we consider a superposition of just two compass states,
continuing our approach of not explicitly normalising states in mathematical expressions but regarding states as implicitly normalised.
We only consider superpositions of compass states comprising superpositions of coherent states with equal magnitudes of displacement from the origin,
i.e., states of the type
$\ket{\diamonds}_{\theta}$~\eqref{eq:compassgeneral}.
For our purposes, only their mutual separation matters. 
Furthermore we restrict to a superpostion of two compass states with mutual rotation~$\nicefrac{\pi}{4}$, namely,
\begin{equation}
\label{eq:twocompass}
\ket{\diamonds}+\ket{\square}
\end{equation}
using the unnormalised-state convention discussed just below Eq.~(\ref{eq:evenoddcatstate}).
We present in this subsection the Wigner function for this state and then the overlap function~$\gamma$.

\begin{figure*}[]
\includegraphics[scale=0.25]{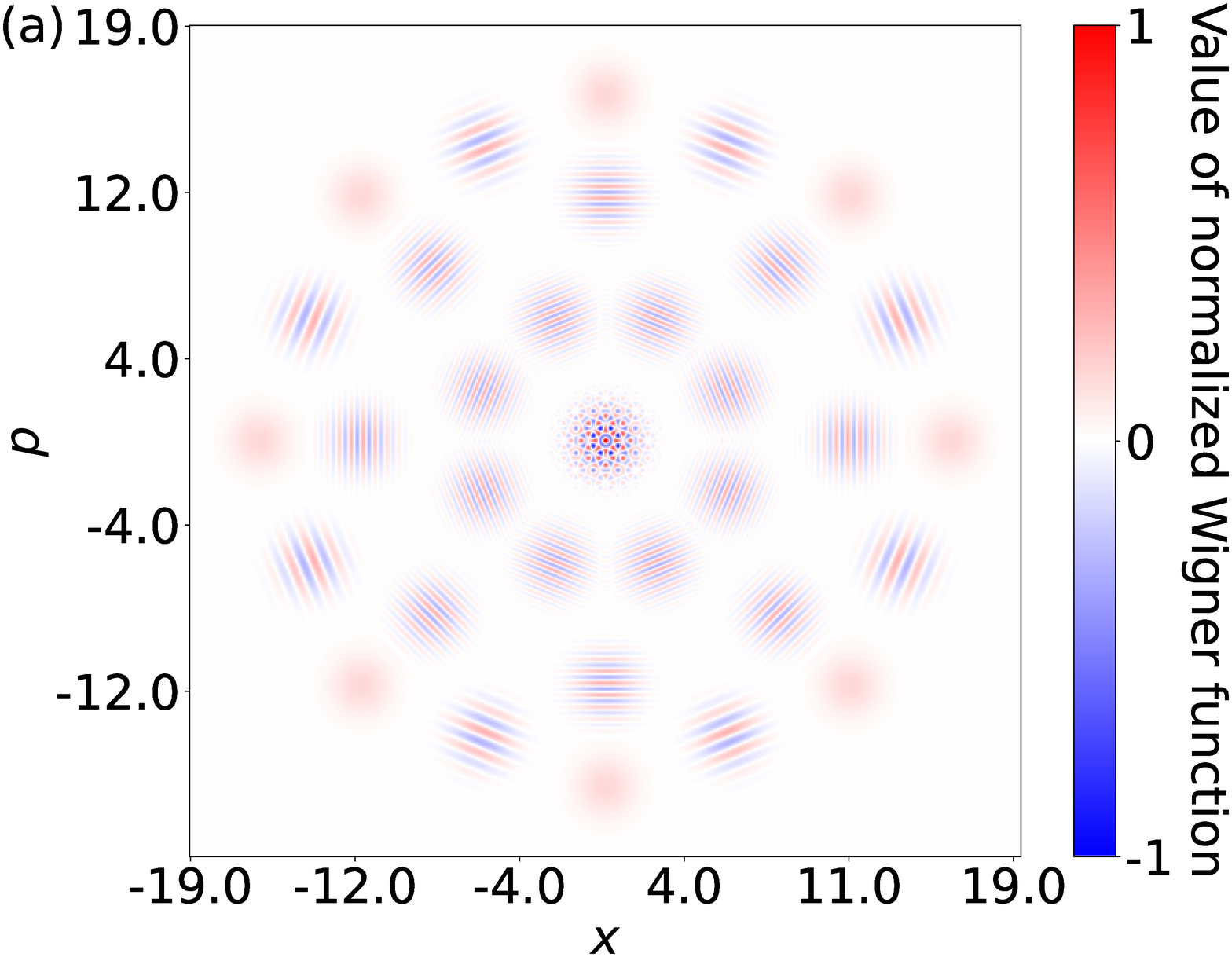}
\includegraphics[scale=0.25]{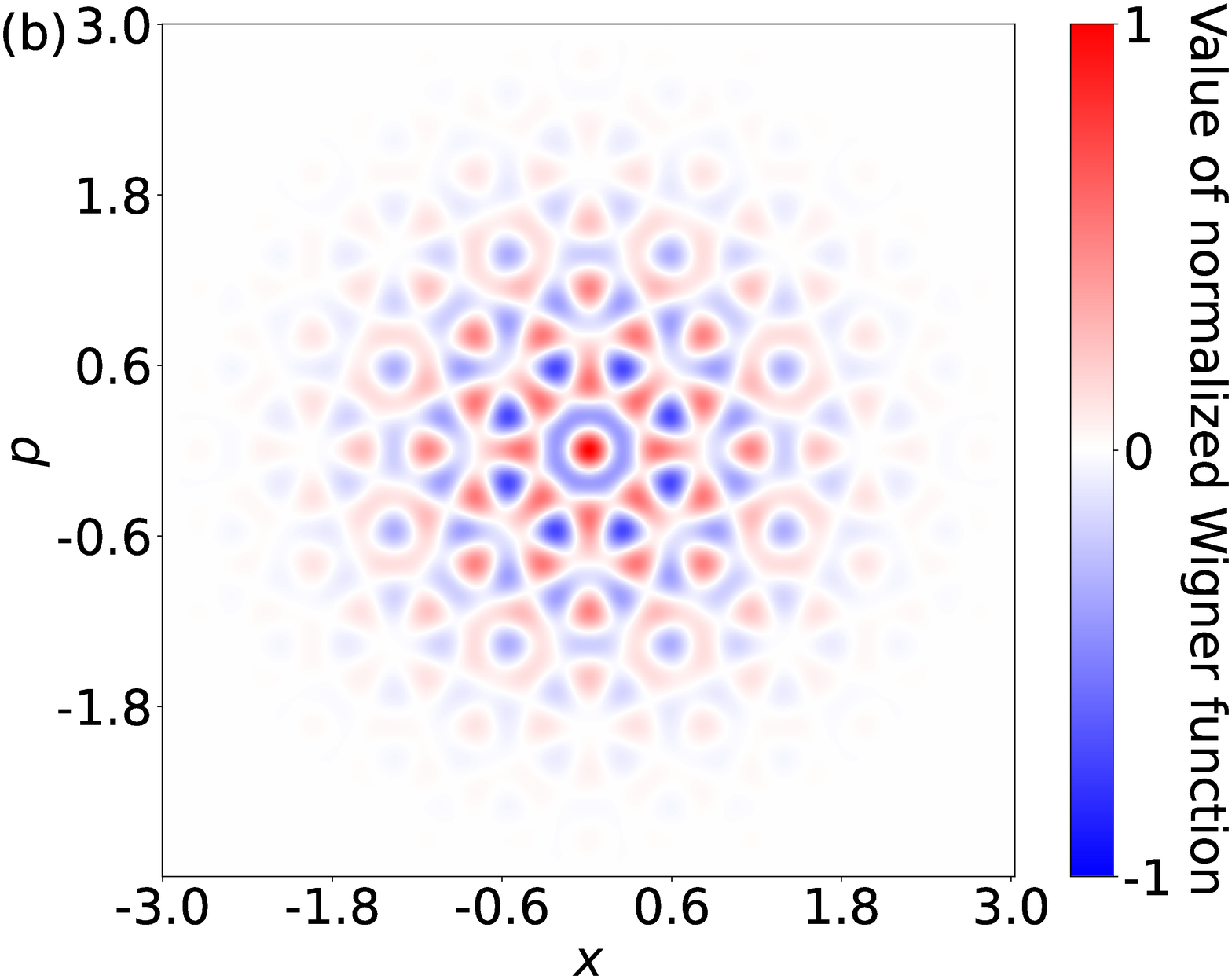}
\caption{%
Heat map for the normalised Wigner function of the superposition of two compass states with $a=8$ (a)~shown fully and (b)~the centre interference pattern only.
}
\label{fig:WignerTwoCompass}
\end{figure*}

\begin{figure*}
\includegraphics[scale = 0.25]{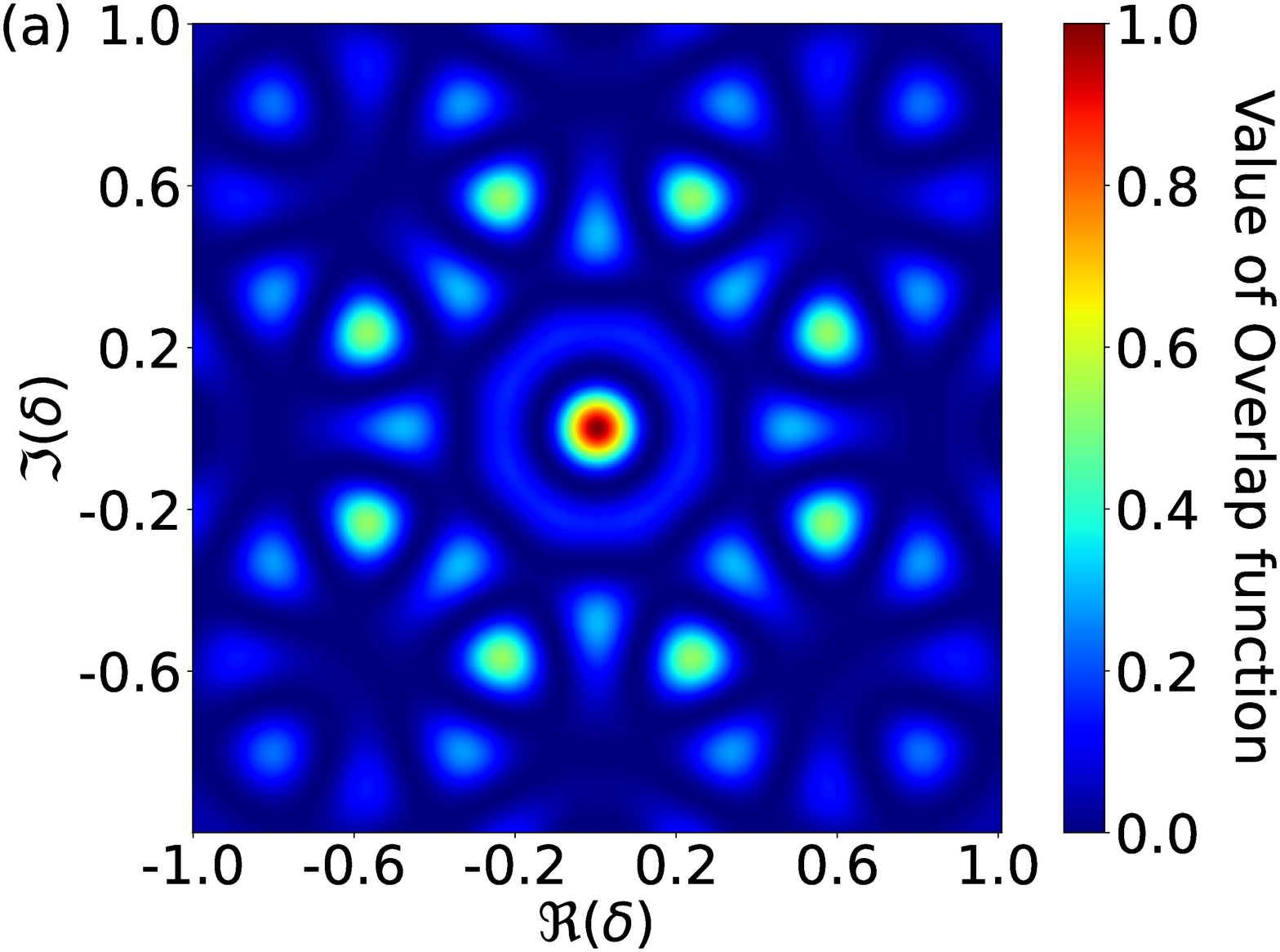}
\includegraphics[scale = 0.25]{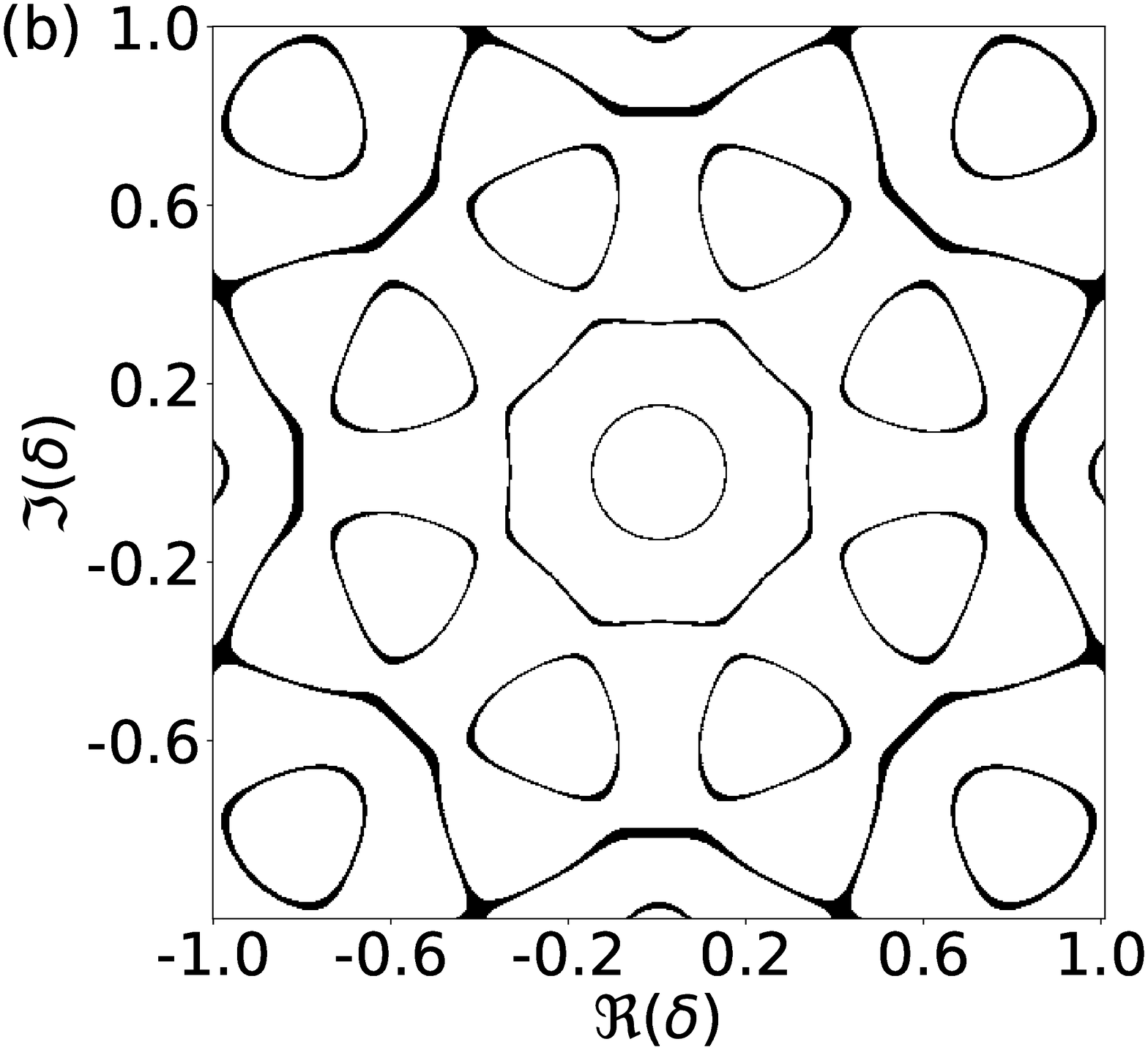}
\caption{Heat map for the overlap function of the superposition of two compass states with $a=8$ (a)~shown fully and (b)~the phase space regions where its value is approximately zero.}
\label{fig:OverlapFunctionTwoCompass}%
\end{figure*}

\paragraph{}
We present a plot of the Wigner function for the superposition of two compass states~\eqref{eq:twocompass} in Fig.~\ref{fig:WignerTwoCompass}. These plots can be compared to the Wigner function plots for a single compass state in Fig.~\ref{fig:WignerOneCompass}. 
The Wigner function consists of a number of rings. The outermost ring corresponds to the coherent states part of our compass states and the interference of the cat state formed by the adjacent coherent states. The next inner ring comprises the interference terms of the cat states formed by alternating coherent states. Following the same pattern, the inner-most ring is the superposition of interference of cat states formed by opposite coherent states. At the centre of the Wigner function, the smallest phase-space structures are found.
\paragraph{}
In the present case of a superposition of two compass states, we have
\begin{equation}
\label{eq:28catstates}
{8\choose2}=28
\end{equation}
cat states.
To calculate the centre interference of the Wigner function, we only consider the cat states formed by the pairs of coherent states maximally separated by angle in phase space as in~\S\ref{sec:one compass}.
The Wigner function, due to the interference at the centre, is

\begin{align}
\label{eq:twocenter}
     W_{\text{cent}}(x,p)\approx& 2\text{e}^{-\nicefrac12(x^2+p^2)}\Big[\cos\left(2xa\right)+\cos\left(2pa\right) \nonumber \\ 
     &+ \cos\left(\sqrt{2}(pa+xa)\right) \nonumber \\ &+
     \cos\left(\sqrt{2}(pa-xa)\right)\Big]. 
\end{align}
For simplicity, we have
only consider the central interference pattern shown in Fig.~\ref{fig:WignerTwoCompass}(b),
as the smallest phase-space structures are at the centre, making central interference pattern the most relevant.
\paragraph{}
For the superposition of two compass states,
we can label coherent states on the circle in phase space by compass coordinates
\begin{equation}
\label{eq:twocompasscoords}
\text{N, NE, E, SE, S, SW, W, NW}
\end{equation}
with NE denoting northeast and so on.
Similarly, 
we write N$'$ etc for coordinates of the eight coherent states of the displaced superposition of two compass states.
Cross terms are denoted by NE-SW$'$ etc.
As we have 8 coherent states,
and cross terms do not include self-terms such as N-N$'$,
$7\times8=56$ cross terms exist.
We do not consider any of these 56 terms;
only the eight self-terms
\begin{align}
\label{eq:8crossterms}
&\text{N-N$'$}, \text{S-S$'$}, \text{E-E$'$}, \text{W-W$'$}, \text{NE-NE$'$}, \text{SW-SW$'$},\nonumber\\
&\text{NW-NW$'$}, \text{SE-SE$'$}
\end{align}
are non-negligible.
Calculating the overlap as in~\S\ref{sec:one compass},
we obtain
\begin{align}
\label{eq:_overlaptwo}
 \gamma(\delta;\ket{\diamonds}+\ket{\square})
 \approx&\text{e}^{-|\delta|^2}
 \Big|\sum_{m=0}^1
\big[\cos\big(2a|\delta|\cos\left(\text{arg}\delta+\nicefrac{m\pi}4\right)\big)\nonumber\\
&+
\cos\big(2a|\delta|\sin\left(\text{arg}\delta+\nicefrac{m\pi}{4}\right)\big)\big]
\Big|^2,
\end{align}
which has a period~$\nicefrac{\pi}4$ over $\arg\delta$. 
\paragraph{}
We plot $\gamma$~\eqref{eq:_overlaptwo} vs $\delta$ and the regions where $\gamma\approx0$ 
by setting a numerical cut-off of
\begin{equation}
\label{eq:numericalcutoff}
\gamma<0.001
\end{equation}
in Fig.~\ref{fig:OverlapFunctionTwoCompass}.
This cut-off is not the threshold $\epsilon$~(\ref{eq:threshold10-15})
but rather a round-off that ensures sufficiently sharp lines in Fig.~\ref{fig:OverlapFunctionTwoCompass}.
The region of phase space where $\gamma\approx0$ encloses the origin. The overlap between the superposition of two compass states and their displaced version goes to zero for any arbitrary direction of displacement.
Now we determine regions in phase space such that the overlap~$\gamma$
is less than the threshold
\begin{equation}
\label{eq:threshold10-15}
    \epsilon\gets10^{-15},
\end{equation}
i.e., assigned the value~$10^{-15}$.
We solve for the roots of Eq.~\eqref{eq:_overlaptwo}, namely,
\begin{align}
\label{eq:overlaptworoots}
\sum_{m=0}^1\big[&\cos\big(2a|\delta|\cos(\text{arg}\delta+\nicefrac{m\pi}{4})\big) \nonumber \\
    &+\cos\big(2a|\delta|\sin(\text{arg}\delta+\nicefrac{m\pi}{4})\big)\big]
    \equiv0,
\end{align}
which is an exact equality but leads only to an approximation of~$\gamma$
 as shown in Eq.~(\ref{eq:_overlaptwo}).
This expression identifies
all~$\delta$ such that~$\gamma(\delta;\ket{\diamonds}+\ket{\square})$ vanishes.
\paragraph{}
To obtain approximate solutions,
we assign the direction of phase-space displacement as~$\text{arg}\delta$ and assign the size~$a$ pertaining to both compass states in the superposition.
We then calculate the Taylor series for Eq.~(\ref{eq:overlaptworoots}) around some $y\in\mathbb{R}$.
Truncating the Taylor series to quadratic order introduces a change of the order of~$10^{-9}$ in the values of~$|\delta|$ for which~$\gamma<\epsilon$.
This truncation is valid because the error due to Taylor series truncation is orders of magnitude smaller than~$\epsilon$.

Now that we know this truncation is valid,
we can neglect cubic and higher-order terms in the expansion, which yields a quadratic expression.
Thus, a good approximation for~$\gamma$ is given by the quadratic equation characterised by the coefficients~$A_2$~(\ref{eq:a_2}), $B_2$~(\ref{eq:b_2}) and~$C_2$~(\ref{eq:c_2})
in the quadratic equation
\begin{equation}
\label{eq:quadratictwocompass}
    A_2|\delta|^2+B_2|\delta|+C_2 = 0
\end{equation}
as elaborated in~$\S\ref{section:simplify_JA_2}$.
The ${}_2$ in the coefficients' subscripts~(\ref{eq:quadratictwocompass})
indicates the number of compass states~$n$ in our superposition.
These coefficients can be 
simplified using the Jacobi-Anger expansion~\eqref{eq:JacobiAnger}
to 
obtain the ultimate expressions for coefficients $A_2$, $B_2$ and $C_2$ of in Eq.~(\ref{eq:quadratictwocompass}).

The coefficients~(\ref{eq:quadratictwocompass}) are valid for the region
\begin{equation}
\label{eq:region}
    \left||\delta| - y\right|\ll~1,
\end{equation}
i.e.,
satisfying the Taylor series truncation approximation,
and direction of phase-space displacement $\text{arg}\delta$.
For~$J_n(z)$ the~$n^\text{th}$ Bessel function of the first kind,
these coefficients are
\begin{align}
\label{eq:a_2_bessel}
     A_2(y;a,\arg\delta)=&4a^2\big[J_2\left(2ay\right)-J_0\left(2ay\right)\big]\nonumber \\ 
                &+2a^2\cos\left(8\text{arg}\delta\right)\big[2J_{6}\left(2ay\right) \nonumber \\
                &+2J_{10}\left(2ay\right)- 4J_{8}\left(2ay\right)\big]
\end{align}
for the quadratic term,
\begin{align}
\label{eq:b_2_bessel}
B_2(y;a,\arg\delta)=&-2A_2y-8aJ_1\left(2ay\right) \nonumber \\
                &+8a\cos\left(8\text{arg}\delta\right)\big[J_{7}(2ay) \nonumber \\
                &-J_{9}\left(2ay\right)\big] 
\end{align}
for the linear term, and
\begin{align} 
\label{eq:c_2_bessel}
    C_2(y;a,\arg\delta)=&-A_2y-B_2+4J_0(2ay) \nonumber \\
    &+8\cos(8\text{arg}\delta)\big[J_{8}(2ay)\big]
\end{align}
for the constant.
All three coefficients have small oscillations with period $\nicefrac{\pi}{4}$ over $\arg\delta$. The roots of Eq.~\eqref{eq:quadratictwocompass} are
\begin{equation}
    |\delta| = \frac{-B_2\pm\sqrt{B_2^2-4A_2C_2}}{2A_2}.
\end{equation}
The roots are the magnitude of displacement $|\delta|$ for which~$\gamma$ vanishes in a specific direction of phase-space displacement~$\text{arg}\delta$.

\paragraph{} 
We now quantify the sensitivity of this state, given by Def.~\ref{def:sensitivity}, for the arbitrary case.
We solve for the roots of Eq.~\eqref{eq:quadratictwocompass}, 
and the resultant coefficients are given by Eqs.~(\ref{eq:a_2_bessel})--(\ref{eq:c_2_bessel}).

In the neighbourhood of~$y=\nicefrac{6}{5a}$, the roots oscillate in the region
\begin{equation}
\label{eq:rootregion}
\frac1a\left[1.20223545,
1.20259051\right]
\end{equation}
with period~$\nicefrac{\pi}4$ over the direction of displacement~$\arg\delta$. The sensitivity of the state is quantified by
\begin{equation}
\label{eq:sensitivitytwocompass}
    |\delta|_{\min} = \nicefrac{1.20223545}{a}    
\end{equation}
which is obtained for
\begin{equation}
\label{eq:sensitivitytwocompassargument}
    \arg\delta = \nicefrac{(2m+1)\pi}{8}.
\end{equation}
As~$|\delta|_\text{min}$ greatly exceeds the Planck scale for~$a\gg1$,
this superposition of two compass states delivers substantial sub-Planck resolution in phase space.

Next, we generalise these results for the case of the superposition of~$n$ compass states.
\subsection{Generalised superposition}
\label{section:generalised_superposition}
We are now in a position to generalise the results to a superposition of $n$ compass states,
each of size $a$
and each rotated by an angle
\begin{equation}
\label{eq:mangle}
\nicefrac{m\pi}{2n},\,
m\in\{0,1,2,\dots,n-1\}.
\end{equation}
Our superposition is thus
\begin{align}
   \sum_{m=0}^{n-1}\ket{\diamonds}_{\nicefrac{m\pi}{2n}}.
\end{align}
As before, we calculate the centre interference part of the Wigner function.
Then we calculate the overlap function for this general case.
Finally, we solve the roots for this overlap to determine sensitivity of this superposition of an arbitrary number of compass states.
\paragraph{}
In the present case of a superposition of~$n$ compass states, we have
\begin{equation}
    {4n\choose2}=\frac{4n(4n-1)}2
\end{equation} 
cat states. To calculate the centre interference of the Wigner function, we only consider the cat states formed by the pairs of coherent states maximally separated by angle,
i.e., by~$\pi$ radians,
in phase space, as done in the previous sections.

The Wigner function, due to the interference at the centre,
is
\begin{align}
\label{eq:ncentre}
W_\text{cen}(x,p)\approx
    &2\text{e}^{-\nicefrac12\left(x^2+p^2\right)}\sum_{m=0}^{n-1}
\bigg[\cos\bigg(2xa\sin\left(\frac{m\pi}{2n}\right)
   \nonumber\\
&-2ap\cos\left(\frac{m\pi}{2n}\right)\bigg)
+\cos\bigg(2ax\sin\left(\frac{m\pi}{2n}\right)\nonumber \\
&+2ap\cos\left(\frac{m\pi}{2n}\right)\bigg)\bigg].
\end{align}
We see a Gaussian envelope as the coefficient for this expression.
This envelope modulates a sum of functions,
and these functions are cosines of a sum of cosine and sine functions
\paragraph{}
For special values of
$n$,
Eq.~(\ref{eq:ncentre}) reduces to our previous Eqs.~(\ref{eq:onecentre}) and~(\ref{eq:twocenter}). The general expression for the Wigner function of the centre interference pattern is given by Eq.~(\ref{eq:ncentre}).
\paragraph{}
We follow the process outlined in the previous sections to neglect cross terms in the overlap. As we have~$4n$ coherent states, and cross terms do not include self-terms such as N-N$'$,
only $4n(4n-1)$ cross terms remain.
We neglect these terms
so only self-terms are non-negligible. Calculating the overlap, we obtain
\begin{align}
\label{eq:overlapn}
 \gamma&\left(\delta; \sum_{m=0}^{n-1}\ket{\diamonds}_{\frac{m\pi}{2n}}\right) \nonumber \\
 &\approx\text{e}^{-|\delta|^2}\Bigg|\sum_{m=0}^{n-1}\bigg[\cos\bigg(2a|\delta|\cos\left(\text{arg}\delta+\frac{m\pi}{2n}\right)\bigg)\nonumber\\
 &+\cos\bigg(2a|\delta|\sin\left(\text{arg}\delta+\frac{m\pi}{2n}\right)\bigg)\bigg]\Bigg|^2
 \end{align}
with a Gaussian coefficient in terms of~$|\delta|^2$
and a sum of cosine functions of sinusoidal functions.
Furthermore,
$\gamma$ exhibits a period of~$\nicefrac{\pi}{2n}$ over~$\arg\delta$ for~$n$ compass states.
We can see that Eq.~\eqref{eq:overlapn} reduces to the special cases of Eqs.~\eqref{eq:overlaponecompass} and
\eqref{eq:_overlaptwo} when~$n\in\{1,2\}$ respectively.
We solve for the roots of Eq.~(\ref{eq:overlapn}), namely,
\begin{align}
    \label{eq:overlapnroots}
    &\sum_{m=0}^{n-1}\big[\cos\big(2a|\delta|\cos\left(\text{arg}\delta+\nicefrac{m\pi}{2n}\right)\big) \nonumber \\
    &+\cos\big(2a|\delta|\sin\left(\text{arg}\delta+\nicefrac{m\pi}{2n}\right)\big)\big]\equiv0,
\end{align}
which is an exact equality but leads only to an approximation of~$\gamma$ as shown in Eq.~\eqref{eq:overlapn}.
Similar to our previous cases,
we truncate the Taylor series to quadratic order to calculate Eqs.~\eqref{eq:general_a}--\eqref{eq:general_c}. As shown in~\S\ref{section:simplify_JA} these equations can be further simplified using Eq.~\eqref{eq:JacobiAnger} to obtain the coefficients~$A_n$,$B_n$ and~$C_n$. The expressions for these coefficients are
\begin{align}
\label{eq:a_n_bessel}
    A_n(y;a,\arg\delta) =   & 2a^2n\big[J_2(2ay)-J_0(2ay)\big]\nonumber \\
                &+na^2\cos\left(4n\arg\delta\right)\big[2J_{4n-2}(2ay)\nonumber \\
                &+2J_{4n+2}(2ay)
                -4J_{4n}(2ay)\big]
\end{align}
for the quadratic term,
\begin{align}
\label{eq:b_n_bessel}
    B_n(y;a,\arg\delta) =  &-2A_ny-4an J_1(2ay) \nonumber \\
                &+4an\cos\left(4n\arg\delta\right)\big[J_{4n-1}(2ay) \nonumber \\
                &-J_{4n+1}(2ay)\big] 
\end{align}
for the linear term, and
\begin{align}
\label{eq:c_n_bessel}
    C_n(y;a,\arg\delta) =  & -A_ny^2-B_ny+2nJ_0(2ay) \nonumber \\
                &+4n\cos\left(4n\arg\delta\right)\big[J_{4n}(2ay)\big]
\end{align}
for the constant.
For the special case of superposition of two compass states $(n=2)$, Eqs.~(\ref{eq:a_n_bessel})--(\ref{eq:c_n_bessel}) reduce to Eqs.~(\ref{eq:a_2_bessel})--(\ref{eq:c_2_bessel}).

Upon increasing the number~$n$ of compass states, the isotropic nature of the superposition of compass state increases because,
as $n\to\infty$,
the Bessel function of the first kind
\begin{equation}
    J_n(y) = \sum_{l=0}^{\infty}\frac{(-1)^l y^{2l+n}}{2^{2l+n}l!(n+l)!}
\end{equation}
tends to zero.
Hence, the oscillatory part for all three coefficients in Eqs.~(\ref{eq:a_n_bessel})--(\ref{eq:c_n_bessel}) vanishes. Due to this, the quadratic equation remains the same for all directions of phase-space displacements. The state thus has uniform sensitivity.

\section{Discussion}
\label{sec:discussions}
Now we give a high-level summary of our results.
To recap,
we have solved for cases of superpositions of a few compass states:
the trivial yet enlightening case of one compass state (to clarify not-so-clear issues in the previous literature),
and superpositions of two compass states.
Then we generalise to the case of a superposition of any finite number of compass states.

In each case, we plot the exact Wigner function and a blown-up version of the centre of the exact Wigner function.
We see these plots for one compass state in Fig.~\ref{fig:WignerOneCompass},
for the superposition of two compass states in Fig.~\ref{fig:WignerTwoCompass}.

In all the cases, we approximate the analytical expression for the centre interference.
We recognise that only the cat states formed by opposite coherent states significantly contribute to our centre interference pattern.
The analytical expressions only consider these terms. This approximation reduces the number of terms significantly as the total terms grow combinatorially with the number of compass states in our superposition.
The plots of centre interference are blown-up versions of our plots of the whole Wigner function.

In all the cases, we calculate the approximate analytical expressions for the overlap functions. We neglect all cross terms that vanish exponentially for increasing~$a$, which previously was implied but not stated explicitly~\cite{Akhtar}.
We plot these approximate overlap functions and the regions where the overlap function is approximately equal to 0 for one compass state in Fig.~\ref{fig:OverlapFunctionOneCompass} and for the superposition of two compass states in Fig.~\ref{fig:OverlapFunctionTwoCompass}. The overlap functions are periodic in all cases. 

The innermost rings in Figs.~\ref{fig:OverlapFunctionOneCompass} and \ref{fig:OverlapFunctionTwoCompass} define the sensitivity. 
The magnitude of the smallest displacement from the origin to a point on this ring is the sensitivity of our state.
As the number of compass states increases, the innermost ring becomes more isotropic in nature; i.e., the magnitude of the oscillations of the ring decreases. This increasing isotropicity can be seen by comparing the case of two compass states and the superposition of three compass states.
The range of oscillation for the periodic inner ring in the case of the superposition of two compass states for an arbitrary value of~$a$ is~$\left[\nicefrac{1.20223545}{a},
\nicefrac{1.20259051}{a}\right]$. 
For the superposition of three compass states this range reduces to~$\left[\nicefrac{1.202412739}{a},
\nicefrac{1.202412805}{a}\right]$. The minimum of this range determines the sensitivity of our superposition. For any superposition, the sensitivity increases as~$a$ increases.
\section{Conclusions}
\label{sec:conclusions}
In conclusion,
we have studied extensively superpositions of compass states,
motivated by extending Zurek's concept of sub-Planck phase-space displacement sensitivity in all directions of phase space~\cite{Zurek2001}
to delivering near-isotropic sub-Planck sensitivity to displacement. The study of sub-Planck phase space sensitivity is important in setting the limits of quantum meters and anticipating the mesh structure required to simulate a quantum system's evolution.
This near-isotropic feature delivers a robustness to ``sub-Planckness'' by not being subject to changes of the phase (weighting of the linear combination of canonical position and momentum).
We attain the ideal case of near-isotropic sensitivity in all phase space displacement directions as the number of compass states in the superposition becomes infinite. This ensures that a perturbation does not favour any phase space direction. A perturbation on such a state will either make the displaced state approximately orthogonal to the original, irrespective of the phase space direction or will not do so in any direction.

We note that the superposition of infinitely many compass states is also a superposition of infinitely many coherent states on a circle in phase space,
which was studied as a basis for representations of any oscillator state~\cite{coherentstatescircles}.
However, our superpositions of an arbitrarily large number of compass states is not a special case of any cases or considerations in that seminal work as their focus on constructing an overcomplete basis focused on designing the phases of coherent states in the superposition.
By analysing superpositions of compass states rather than this more general case, near-isotropic sub-Planckness in phase space is clearer and even intuitive although, as we have shown, the mathematics is somewhat complicated.

Future work could consider experimental realisations of superpositions of compass states.
Experimental realisations of a compass state have been explored~\cite{Praxmeyer,Vlastakis,Ofek2016ExtendingTL},
and an experimental realisation could follow in that vein.
These superpositions of compass states could have an impact on ultra-sensitive quantum metrology and efficient storage of quantum information, similar to the suggested impact of one compass state alone.

\acknowledgments
We acknowledge the traditional owners of the land on which this work was undertaken at the University of Calgary:
the Treaty 7 First Nations (www.treaty7.org).
This project was supported by MITACS and NSERC.
\appendix
\label{sec:appendix}
\section{Superposition of three compass states}
\label{sec:three compass}
\paragraph{}
Now we consider a superposition of just three compass states,
continuing our approach of not explicitly normalising states in mathematical expressions but regarding states as implicitly normalised. Similar to the case of the superposition of two compass states, we consider coherent states with equal magnitudes~$a$ of displacement from the origin in our superposition.
We restrict to a superposition of three compass states with mutual rotation~$\nicefrac{\pi}{6}$, namely,

\begin{align}
\label{eq:threecompass}
&\ket{\square}+\ket{{\rotatebox[origin=c]{30}{$\square$}}}
+\ket{{\rotatebox[origin=c]{60}{$\square$}}},\, \nonumber \\
&\ket{{\rotatebox[origin=c]{30}{$\square$}}}
:=\ket{\square}_{30^\circ},\, \ket{{\rotatebox[origin=c]{60}{$\square$}}} 
:=\ket{\square}_{60^\circ}.
\end{align}
We present in this subsection the Wigner function for this state and then the overlap function~$\gamma$.
\paragraph{}
Now we calculate and plot the Wigner function for the superposition of three compass states~(\ref{eq:threecompass}) using approximations similar to those in~\S\ref{sec:two compass}.
Specifically,
we plot the entire and exact Wigner function.
Then we explain how we approximate the centre interference pattern for the Wigner function.

The Wigner function is similar to the previous case of the superposition of two compass states and consists of multiple rings. These rings correspond to different interference terms of the cat states formed by the constituent coherent states of our compasses as we have explained before.
\begin{figure*}
\includegraphics[scale = 0.25]{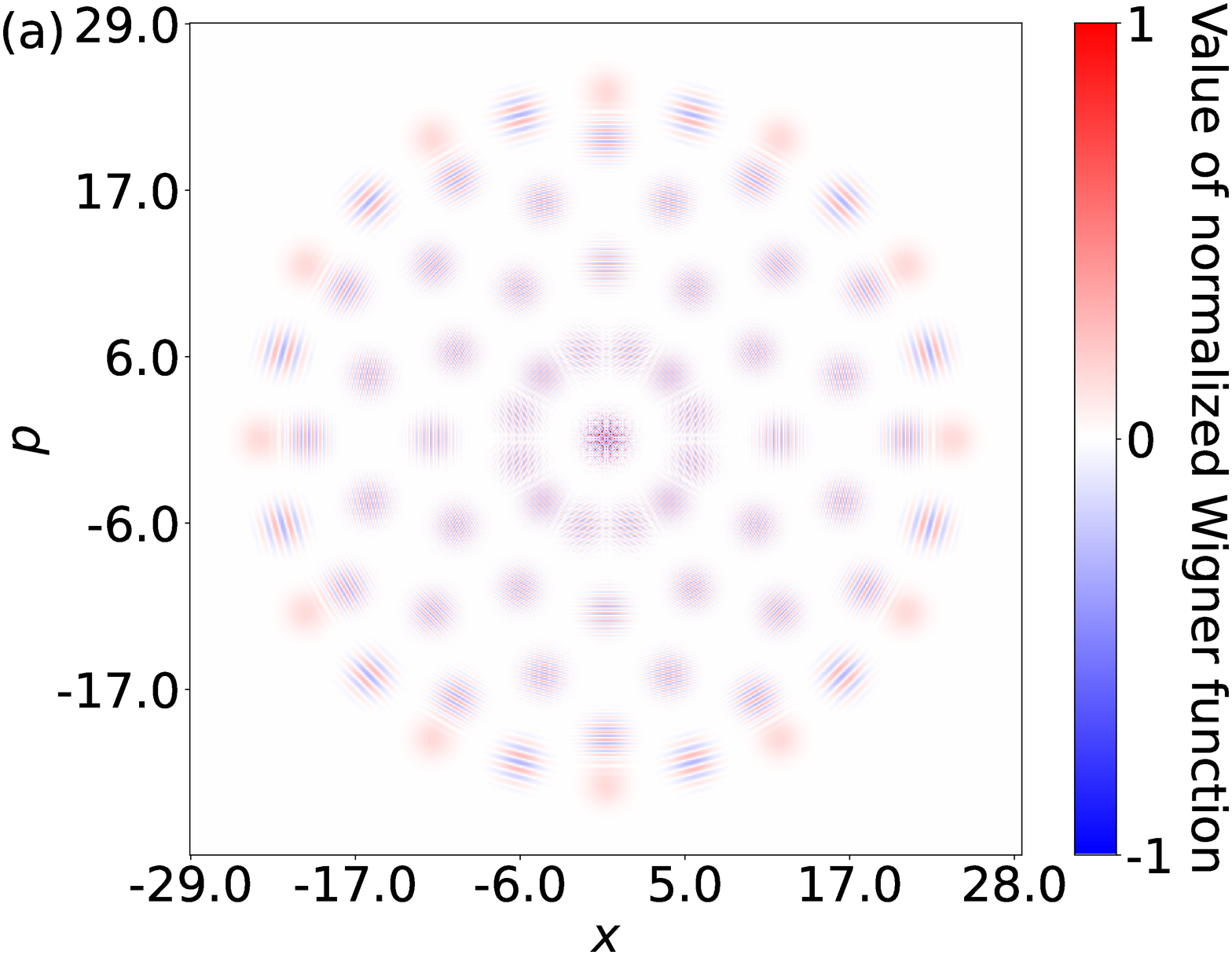}
\includegraphics[scale = 0.25]{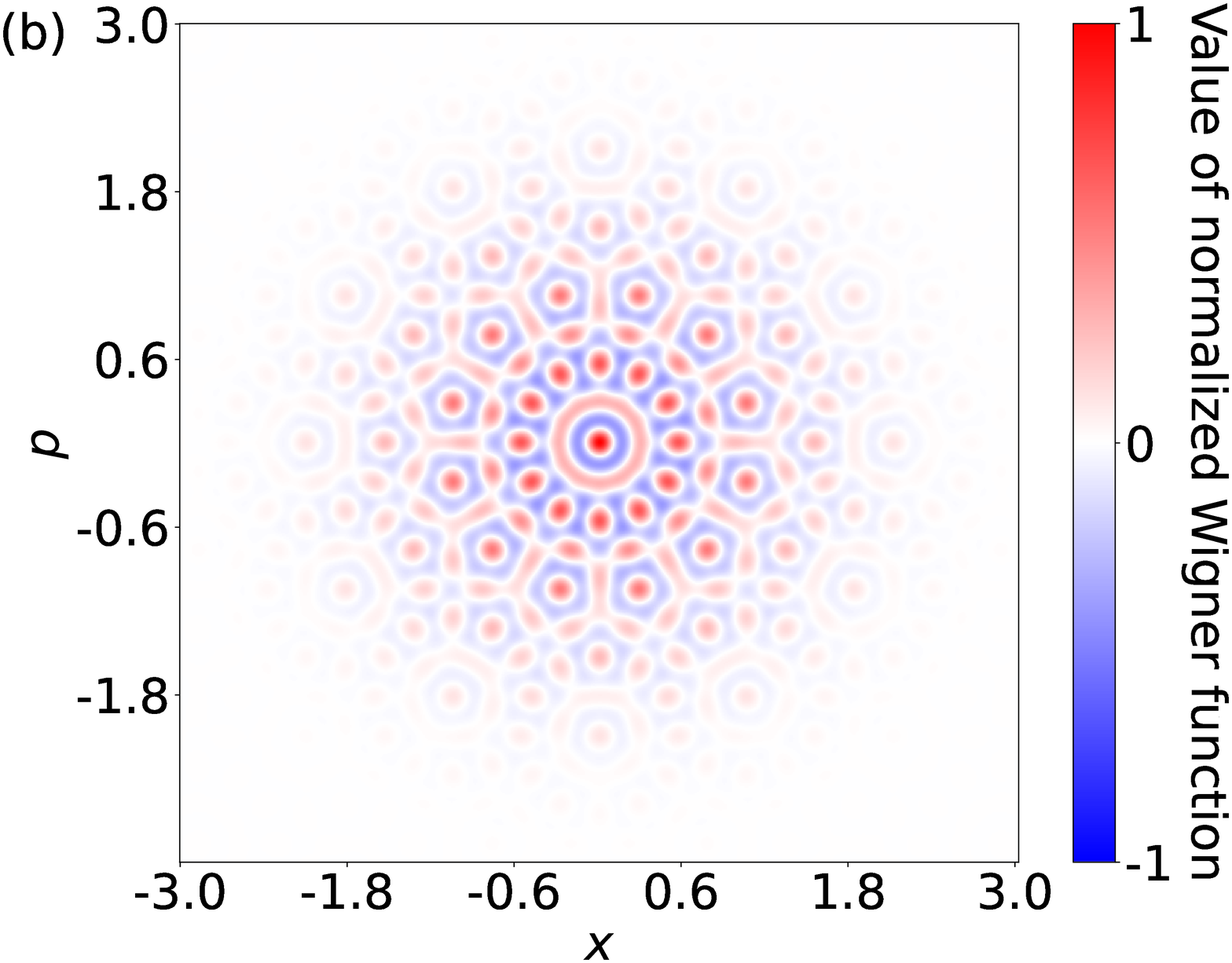}
\caption{%
Heat map for the normalised Wigner function of the superposition of three compass states with $a=12$ (a)~shown fully and (b)~the centre interference pattern only.
}
\label{fig:WignerThreeCompass}
\end{figure*}

\begin{figure*}
\includegraphics[scale = 0.25]{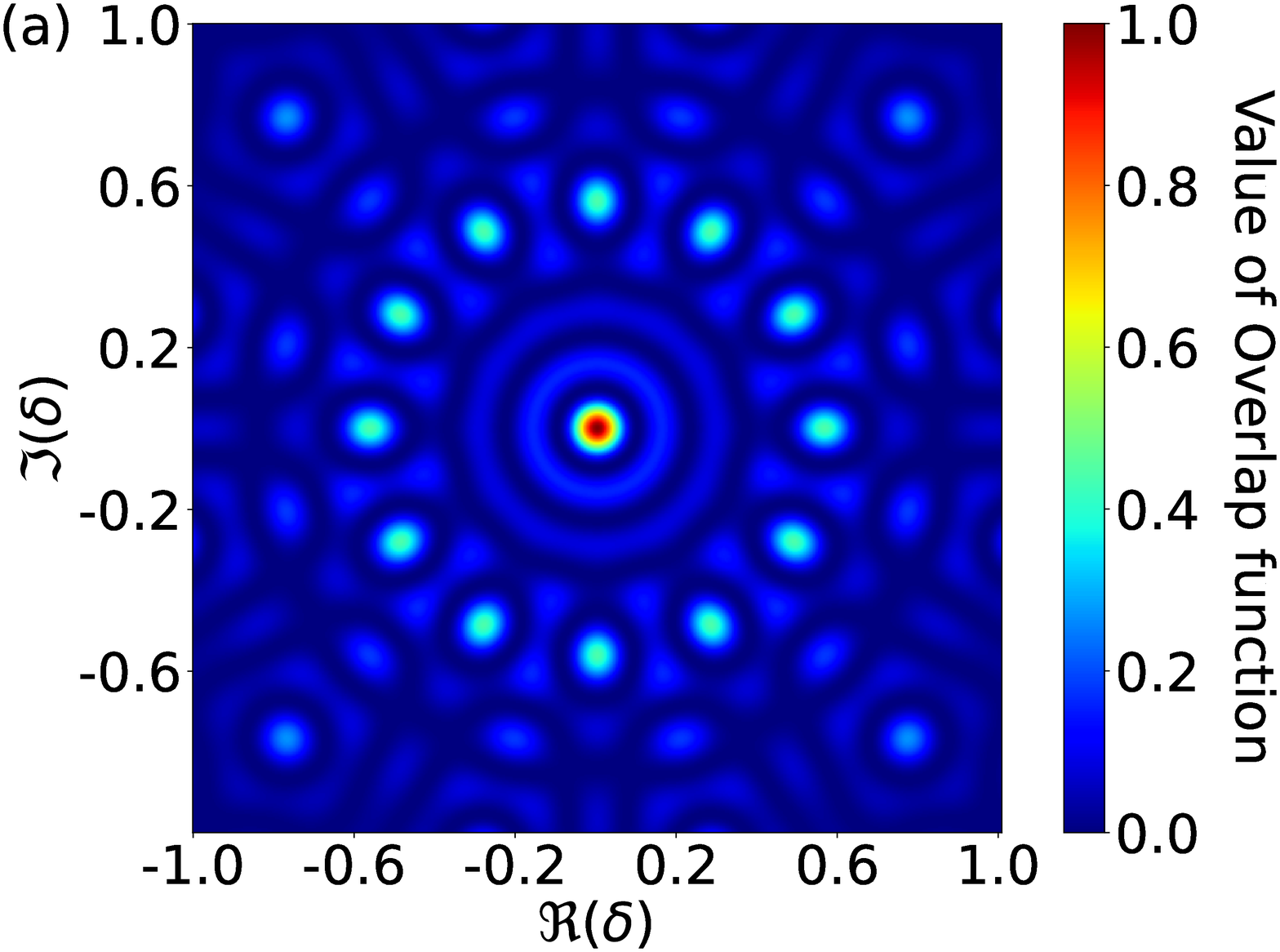}
\includegraphics[scale = 0.25]{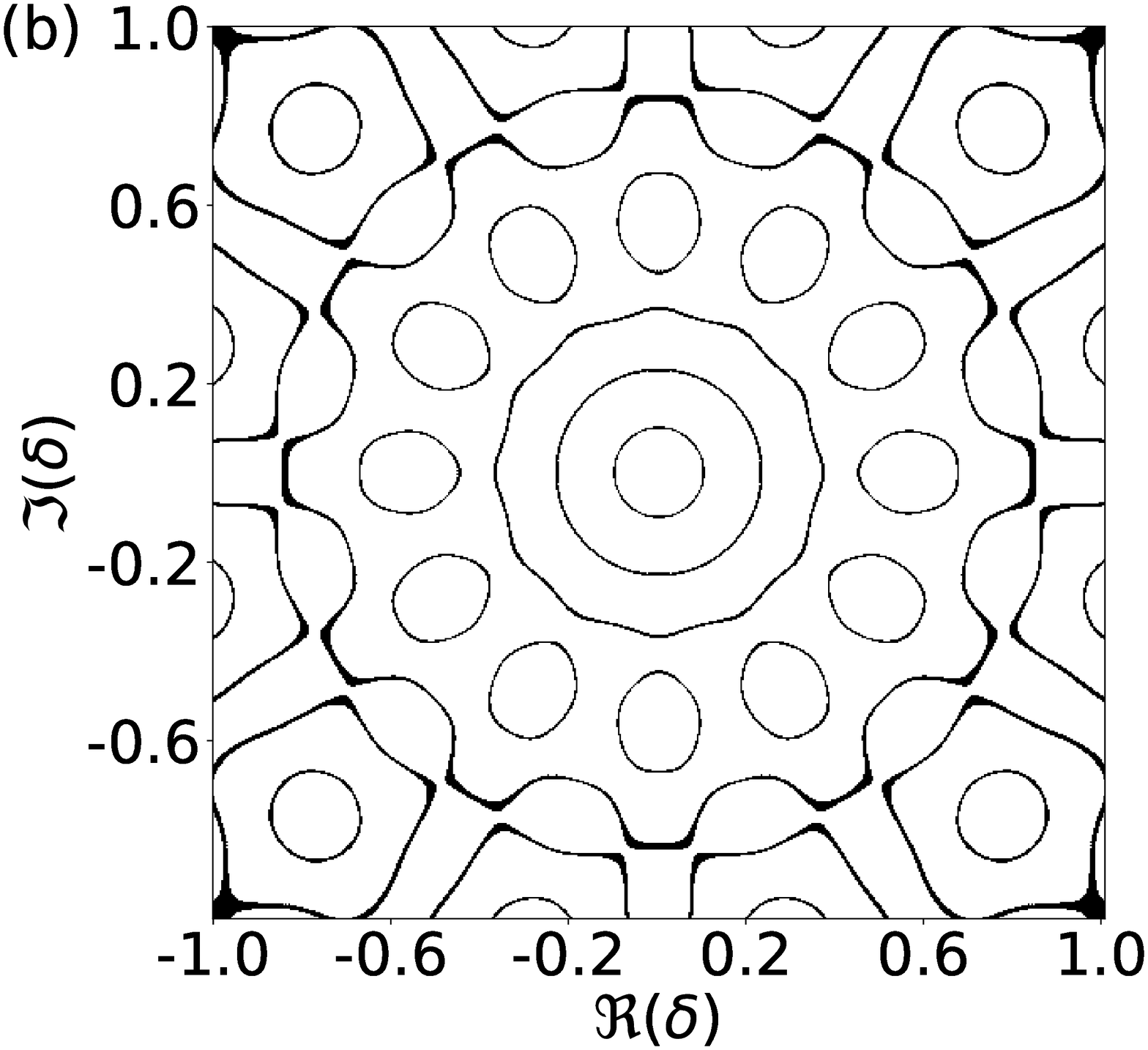}
\caption{Heat map for the overlap function of the superposition of three compass states with $a=12$ (a)~shown fully and (b)~the phase space regions where its value is approximately zero.}
\label{fig:OverlapFunctionThreeCompass}%
\end{figure*}

\paragraph{}
Now we focus on the centre interference pattern Fig.~(\ref{fig:WignerThreeCompass}(b).
In the present case of a superposition of three compass states, we have
\begin{equation}
\label{eq:66catstates}
{12\choose2}=66
\end{equation}
cat states. 
To calculate the centre interference of the Wigner function, we only consider the cat states formed by the pairs of coherent states maximally separated by angle,
i.e., $\pi$ radians,
in phase space as in~\S\ref{sec:one compass}.
The Wigner function at the centre is
\begin{align}
\label{eq:threecentre}
     W_\text{cen}(x,p)
     \approx& 2\text{e}^{-\nicefrac12(x^2+p^2)}
     \Big[\cos\left(2xa\right)+\cos\left(2pa\right)\nonumber\\
     &+\cos\left(\sqrt3xa+pa\right)+
     \cos\left(\sqrt3pa-xa\right)\nonumber\\
     &+\cos\left(\sqrt3pa+xa\right)\nonumber \\
     &+\cos\left(\sqrt3xa-pa\right)\Big]
\end{align}
from six maximally separated pairs of coherent states (cats).
Next, we focus on the overlap function.
We follow the process outlined in~\S\ref{sec:two compass} to neglect the cross terms in the overlap.
As we have 12 coherent states,
i.e., 4 coherent states per compass state and three compass states,
and,
as cross terms do not include self-terms such as N-N$'$,
precisely $11\times12=132$ cross terms exist.
We do not consider any of these 132 terms;
only the twelve self-terms are non-negligible.
Calculating the overlap as in~\S\ref{sec:one compass},
we obtain
\begin{align}
\label{eq:overlapthree}
\gamma&\left(\delta;\ket{\square}+\ket{\text{\rotatebox[origin=c]{30}{$\square$}}}
+\ket{\text{\rotatebox[origin=c]{60}{$\square$}}}\right)\nonumber\\
&\approx\text{e}^{-|\delta|^2}\Big|\sum_{m=0}^2\big[\cos\big(2a|\delta|\cos\left(\text{arg}\delta+\nicefrac{m\pi}{6}\right)\big)\nonumber \\
&+\cos\big(2a|\delta|\sin\left(\text{arg}\delta+\nicefrac{m\pi}{6}\right)\big)\big]\Big|^2,
\end{align}
which has a period of $\nicefrac{\pi}{6}$ over $\arg\delta$.
\paragraph{}
We plot $\gamma$~\eqref{eq:overlapthree} vs $\delta$ and the regions where $\gamma\approx0$ by setting a numerical cut-off of
$\gamma<0.001$ in Fig.~\ref{fig:OverlapFunctionThreeCompass}. 
This cut-off is not the threshold~$\epsilon$.
The overlap between the superposition of three compass states and its displaced version goes to zero for any arbitrary direction of displacement. 

Now we determine regions in phase space such that the overlap~$\gamma$
is less than the threshold~$\epsilon$,
which we set as $10^{-15}$~(\ref{eq:threshold10-15}).
We solve for the roots of Eq.~(\ref{eq:overlapthree}), namely,
\begin{align}
    \label{eq:overlapthreeroots}
&\sum_{m=0}^2\big[\cos\big(2a|\delta|\cos\left(\text{arg}\delta+\nicefrac{m\pi}{6}\right)\big) \nonumber \\
    &+\cos\big(2a|\delta|\sin\left(\text{arg}\delta+\nicefrac{m\pi}{6}\right)\big)\big]\equiv0,
\end{align}
which is an exact equality but leads only to an approximation of~$\gamma$ as shown in Eq.~\eqref{eq:overlapthree}.
Equation~(\ref{eq:overlapthreeroots}) identifies all~$\delta$ such that
\begin{equation}
\gamma(\delta;\ket{\square}+\ket{\text{\rotatebox[origin=c]{30}{$\square$}}}
+\ket{\text{\rotatebox[origin=c]{60}{$\square$}}})
\end{equation}
vanishes.
\paragraph{}
We follow the process outlined in~\S\ref{sec:two compass} and truncate the Taylor series to quadratic order. This introduces a change of the order of~$10^{-9}$ in the values of~$|\delta|$ for which~$\gamma<\epsilon$.
\paragraph{}
The final expressions for the coefficients of the quadratic 
equation
\begin{equation}
\label{eq:quadraticthreecompass}
A_3(|\delta|)^2+B_3(|\delta|)+C_3 = 0
\end{equation}
for a fixed $\arg\delta$ and $a$ for the region~(\ref{eq:region}) are
\begin{align}
\label{eq:a_3_bessel}
     A_3(y;a,\arg\delta)=&6a^2\big[J_2\left(2ay\right)-J_0\left(2ay\right)\big]\nonumber \\ 
                &+3a^2\cos\left(12\arg\delta\right) \nonumber
                \\
                &\times\big[2J_{10}\left(2ay\right)+2J_{14}\left(2ay\right) \nonumber \\
                &- 4J_{12}\left(2ay\right)\big] 
\end{align}
for the quadratic term,
\begin{align}
\label{eq:b_3_bessel}
     B_3(y;a,\arg\delta)=&-2A_3y-12aJ_1\left(2ay\right)\nonumber \\
                &+12a\cos\left(12\arg\delta\right)\big[J_{11}\left(2ay\right) \nonumber \\
                &-J_{13}\left(2ay\right)\big] 
\end{align}
for the linear term, and
\begin{align}
\label{eq:c_3_bessel}
     C_3(y;a,\arg\delta)=&-A_3y-B_3+6J_0\left(2ay\right) \nonumber \\ 
                &+12\cos\left(12\arg\delta\right)\big[J_{12}(2ay)\big]
\end{align}
for the constant.
Each coefficient comprises a constant and an oscillatory term similar to the case of the superposition of two compass states. The oscillations are periodic with period~$\nicefrac{\pi}{6}$ over~$\arg\delta$. The roots of Eq.~\eqref{eq:quadraticthreecompass} are
\begin{equation}
    |\delta| = \frac{-B_3\pm\sqrt{B_3^2-4A_3C_3}}{2A_3}
\end{equation}
with an angular period of~$\nicefrac{\pi}6$
evident in Fig.~\ref{fig:OverlapFunctionThreeCompass}(b)
\paragraph{}
We now quantify the sensitivity of this state, given by
Def.~\ref{def:sensitivity} for the arbitrary case when the size pertaining
to the three compass states of our superposition is~$a$.
We solve the roots of Eq.~\eqref{eq:quadraticthreecompass} characterised by Eqs.~(\ref{eq:a_3_bessel})--(\ref{eq:c_3_bessel}) for different values of $\arg\delta$.

In the neighbourhood of~$y=\nicefrac{6}{5a}$, the roots oscillate in the region
\begin{equation}
\label{eq:rootregion2}
\frac1a\left[1.202412739,
1.202412805\right]
\end{equation}
with period~$\nicefrac{\pi}6$ over the direction of displacement~$\arg\delta$.
The sensitivity of the state is quantified by
\begin{equation}
\label{eq:sensitivitythreecompass}
|\delta|_\text{min}=\frac{1.202412739}{a},    
\end{equation}
which is obtained for
\begin{equation}
\label{eq:sensitivitythreecompassargument}
\arg\delta = \frac{(2m+1)\pi}{12}.
\end{equation}
The amplitude of oscillation has reduced when compared with the case of the superposition of two compass states;
i.e., the isotropicity has increased with respect to the superposition of two compass states.

\begin{widetext}
\section{Exact overlap function of one compass state}
In this section, we provide the exact overlap for the case of one compass state. In the following expression, we also consider the cross terms. These cross-terms correspond to the 
\begin{itemize}
    \item north-south$'$ (NS$'$), north-east$'$ (NE$'$), north-west$'$ (NW$'$),
    \item south-north$'$ (SN$'$), south-east$'$ (SE$'$), south-west$'$ (SW$'$),
    \item east-north$'$ (EN$'$), east-west$'$ (EW$'$), east-south$'$ (ES$'$),
    \item west-north$'$ (WN$'$), west-east$'$ (WE$'$) and west-south$'$ (WS$'$)
\end{itemize}
cases. The exact expression is given by
\begin{align}
\label{eq:exactoverlap}
    \left|\bra{\diamonds}D(\delta)\ket{\diamonds}\right|^2 =& \Big| 2\text{e}^{-\nicefrac{|\delta|^2}{2}}\cos(2a\text{Im}(\delta))+2\text{e}^{-\nicefrac{|\delta|^2}{2}}\cos(2a\text{Re}(\delta))+\text{e}^{\nicefrac{-|2a-\delta|^2}{2}}+\text{e}^{\text{i}(a\text{Im}(\delta)-a\text{Re}(\delta)+a^2)-\nicefrac{|a-\text{i}a-\delta|^2}{2}} \nonumber \\
    & +\text{e}^{\text{i}(a\text{Im}(\delta)+a\text{Re}(\delta)-a^2)-\nicefrac{|a+\text{i}a-\delta|^2}{2}}+\text{e}^{\text{i}(-a\text{Im}(\delta)-a\text{Re}(\delta)-a^2)-\nicefrac{|-a-\text{i}a-\delta|^2}{2}} \nonumber \\
    & +\text{e}^{\text{i}(-a\text{Im}(\delta)+a\text{Re}(\delta)+a^2)-\nicefrac{|-a+\text{i}a-\delta|^2}{2}} \nonumber \\
    & +\text{e}^{\text{i}(a\text{Im}(\delta)-a\text{Re}(\delta)-a^2)-\nicefrac{|a+\text{i}a-\delta|^2}{2}}+\text{e}^{\text{i}(-a\text{Im}(\delta)-a\text{Re}(\delta)+a^2)-\nicefrac{|a+\text{i}a-\delta|^2}{2}} \nonumber \\
    & +\text{e}^{\text{i}(a\text{Im}(\delta)+a\text{Re}(\delta)+a^2)-\nicefrac{|-a-\text{i}a-\delta|^2}{2}}+\text{e}^{\text{i}(-a\text{Im}(\delta)+a\text{Re}(\delta)-a^2)-\nicefrac{|a-\text{i}a-\delta|^2}{2}}+\text{e}^{\nicefrac{-|-2a-\delta|^2}{2}} \nonumber \\
    & +\text{e}^{\nicefrac{-|-2\text{i}a-\delta|^2}{2}}+\text{e}^{\nicefrac{-|-2\text{i}a-\delta|^2}{2}}\Big|^2.
\end{align}
Making the radius of the circle, on which the coherent states are placed, significantly larger than Planck scale $a=1$
makes the cross-terms vanish exponentially.
\section{Details of Taylor Series expansion for the superposition of two compass states}
\label{section:taylorSeriesdetails_2}
For the specific case of the superposition of two compass states, the overlap~\eqref{eq:_overlaptwo} vanishes when,
\begin{align}
\label{eq:appoverlaptwo}
    & \sum_{m=0}^{1}\Big[\cos\big(2a|\delta|\cos\left(\text{arg}\delta+\nicefrac{m\pi}{4}\right)\big) +\cos\big(2a|\delta|\sin\left(\text{arg}\delta+\nicefrac{m\pi}{4}\right)\big)\Big]\equiv0.
\end{align}
To obtain approximate solutions, we assign the direction of phase-space displacement as~$\text{arg}\delta$ and assign the size~$a$ pertaining to both compass states in the superposition. We then calculate the Taylor series for Eq.~\eqref{eq:appoverlaptwo} around some~$y\in\mathbb{R}$. We truncate the Taylor series to quadratic order. The Taylor series of the first term in Eq.~\eqref{eq:appoverlaptwo} is
\begin{align}
\cos\big(2a|\delta|\cos\left(\arg\delta+\nicefrac{m\pi}{4}\right)\big) = &\cos\big(2ay\cos\left(\arg\delta+\nicefrac{m\pi}{4}\right)\big)-\sin\big(2ay\cos\left(\arg\delta+\nicefrac{m\pi}{4}\right)\big)2a\cos\left(\arg\delta+\nicefrac{m\pi}{4}\right)\left(|\delta|-y\right) \nonumber \\
&-\cos\big(2ay\cos\left(\arg\delta+\nicefrac{m\pi}{4}\right)\big)2a^2\cos^2\left(\arg\delta+\nicefrac{m\pi}{4}\right)\left(|\delta|-y\right)^2 
\end{align}
and the Taylor series of the second term in Eq.~\eqref{eq:appoverlaptwo} is
\begin{align}
\cos\big(2a|\delta|\sin\left(\arg\delta+\nicefrac{m\pi}{4}\right)\big) = &\cos\big(2ay\sin\left(\arg\delta+\nicefrac{m\pi}{4}\right)\big)-\sin\big(2ay\sin\left(\arg\delta+\nicefrac{m\pi}{4}\right)\big)2a\sin\left(\arg\delta+\nicefrac{m\pi}{4}\right)(|\delta|-y) \nonumber \\
&-\cos\big(2ay\sin\left(\arg\delta+\nicefrac{m\pi}{4}\right)\big)2a^2\sin^2\left(\arg\delta+\nicefrac{m\pi}{4}\right)(|\delta|-y)^2.
\end{align}
The Taylor series of Eq.~\eqref{eq:appoverlaptwo} is the sum of both these series.
\section{Form of quadratic coefficients for the superposition of two compass states}
\label{section:simplify_JA_2}
The coefficients of the quadratic equation~$(A_2(|\delta|)^2+B_2(|\delta|)+C_2)$ for a fixed~$\arg\delta$ and~$a$ for the region~$||\delta|-y|\ll1$ are
\begin{align}
\label{eq:a_2}
     A_2(y;a,\arg\delta) = &\sum_{m=0}^{1}\Big[-2a^2\sin^2\left(\arg\delta+\nicefrac{m\pi}{4}\right) \cos\big(2ay\sin\left(\arg\delta+\nicefrac{m\pi}{4}\right)\big) \nonumber \\
     &-2a^2\cos^2\left(\arg\delta+\nicefrac{m\pi}{4}\right) \cos\big(2ay\cos\left(\arg\delta+\nicefrac{m\pi}{4}\right)\big)\Big] 
\end{align}
for the quadratic term,
\begin{align}
\label{eq:b_2}
     B_2(y;a,\arg\delta) = &\sum_{m=0}^{1} \Big[4a^2y\sin^2\left(\arg\delta+\nicefrac{m\pi}{4}\right)\cos\big(2ay\sin\left(\arg\delta+\nicefrac{m\pi}{4}\right)\big) \nonumber \\
     &-2a\sin\left(\arg\delta+\nicefrac{m\pi}{4}\right)\sin\big(2ay\sin\left(\arg\delta+\nicefrac{m\pi}{4}\right)\big)  \nonumber \\&
    + 4a^2y\cos^2\left(\arg\delta+\nicefrac{m\pi}{4}\right)\cos
    \big(2ay\cos\left(\arg\delta+\nicefrac{m\pi}{4}\right)\big) \nonumber \\
    &-2a\cos\left(\arg\delta+\nicefrac{m\pi}{4}\right)\sin\big(2ay\cos\left(\arg\delta+\nicefrac{m\pi}{4}\right)\big)\Big] 
\end{align}
for the linear term, and
\begin{align}
\label{eq:c_2}
 C_2(y;a,\arg\delta) =  &\sum_{m=0}^{1}
         \Big[2ay\sin\left(\arg\delta+\nicefrac{m\pi}{4}\right)\sin\big(2ay\sin\left(\arg\delta+\nicefrac{m\pi}{4}\right)\big) \nonumber \\
         &-2a^2y^2\sin^2\left(\arg\delta+\nicefrac{m\pi}{4}\right) \cos\big(2ay\sin\left(\arg\delta+\nicefrac{m\pi}{4}\right)\big)  +\cos\big(2ay\sin\left(\arg\delta+\nicefrac{m\pi}{4}\right)\big) \nonumber \\ &+ \cos\big(2ay\cos\left(\arg\delta+\nicefrac{m\pi}{4}\right)\big) 
         + 2ay\cos\left(\arg\delta+\nicefrac{m\pi}{4}\right)\sin\big(2ay\cos\left(\arg\delta+\nicefrac{m\pi}{4}\right)\big)\nonumber \\ &-2a^2y^2\cos^2\left(\arg\delta+\nicefrac{m\pi}{4}\right) \cos\big(2ay\cos\left(\arg\delta+\nicefrac{m\pi}{4}\right)\big)\Big].
\end{align}
for the constant.
We can simplify these to obtain Eqs.~(\ref{eq:a_2_bessel})--(\ref{eq:c_2_bessel}) using the Jacobi-Anger expansions of $\cos(z\cos\theta)$, $\cos(z\sin\theta)$, $\sin(z\sin\theta)$ and $\sin(z\cos\theta)$ given by
\begin{align}
\label{eq:JacobiAnger}
    &\cos(z\cos\theta) = J_0(z)+2\sum_{n=1}^{\infty} (-1)^n J_{2n}(z)\cos(2n\theta), \nonumber \\
    &\cos(z\sin\theta) 
    = J_0(z)+2\sum_{n=1}^{\infty}  J_{2n}(z)\cos(2n\theta), \nonumber \\ 
  &\sin(z\sin\theta) = 2\sum_{n=1}^{\infty}  J_{2n-1}(z)\sin\big[(2n-1)\theta\big], \nonumber\\ 
  &\sin(z\cos\theta) = -2\sum_{n=1}^{\infty} (-1)^n J_{2n-1}(z)\cos\big[(2n-1)\theta\big].
\end{align}
We further simplify Eq.~(\ref{eq:a_2}) to
\begin{align}
\label{eq:intermediateA2}
        A_2(y;a,\arg\delta) = &\sum_{m=0}^{1} \Big[-2a^2\sin^2\left(\arg\delta+\nicefrac{m\pi}{4}\right) \cos\big(2ay\sin\left(\arg\delta+\nicefrac{m\pi}{4}\right)\big) \nonumber \\
     &-2a^2\cos^2\left(\arg\delta+\nicefrac{m\pi}{4}\right) \cos\big(2ay\cos\left(\arg\delta+\nicefrac{m\pi}{4}\right)\big)\Big] \nonumber \\
              =&\sum_{m=0}^{1}\Big[-a^2\big(1-\cos2(\arg\delta+\nicefrac{m\pi}{4})\big)\cos\big(2ay\sin(\arg\delta+\nicefrac{m\pi}{4})\big) \nonumber \\
              &-a^2\big(1+\cos2(\arg\delta+\nicefrac{m\pi}{4})\big)\cos\left(2ay\cos(\arg\delta+\nicefrac{m\pi}{4})\right)\Big] \nonumber \\
              =&\sum_{m=0}^{1}-\Big[a^2\Big(\cos\big(2ay\sin\left(\arg\delta+\nicefrac{m\pi}{4}\right)\big)+\cos\big(2ay\cos\left(\arg\delta+\nicefrac{m\pi}{4}\right)\big)\Big) \nonumber \\
              &+a^2\cos2(\arg\delta+\nicefrac{m\pi}{4})\Big(\cos\big(2ay\sin\left(\arg\delta+\nicefrac{m\pi}{4}\right)\big)-\cos\big(2ay\cos(\arg\delta+\nicefrac{m\pi}{4})\big)\Big)\Big]. 
\end{align}
\par
Further simplifying with the help of Eq.~\eqref{eq:JacobiAnger}, we obtain
\begin{align}
    \cos\big(2ay\sin\left(\arg\delta+\nicefrac{m\pi}{4}\right)\big)+\cos\big(2ay\cos\left(\arg\delta+\nicefrac{m\pi}{4}\right)\big) =& 2J_0(2ay)+4\sum_{s=1}^{\infty}J_{4s}\left(2ay\right)\cos\big(4s\left(\arg\delta+\nicefrac{m\pi}{4}\right)\big) 
\end{align}
and
\begin{align}
    \cos\big(2ay\sin\left(\arg\delta+\nicefrac{m\pi}{4}\right)\big)-\cos\big(2ay\cos\left(\arg\delta+\nicefrac{m\pi}{4}\right)\big) =& 4\sum_{s=1}^{\infty}J_{4s-2}(2ay)\cos\big(\left(4s-2\right)\left(\arg\delta+\nicefrac{m\pi}{4}\right)\big).
\end{align}
Inserting these back in Eq.~\eqref{eq:intermediateA2} in yields
\begin{align}
    A_2(y;a,\arg\delta)=&4a^2J_2(2ay)-4a^2J_0(2ay)+\sum_{m=0}^{1}\sum_{s=1}^{\infty}\Big[\big(2J_{4s-2}(2ay)+2J_{4s+2}(2ay)-4J_{4s}(2ay)\big)a^2\cos\big(4s\left(\arg\delta+\nicefrac{2\pi m s}{2}\right)\big)\Big] \nonumber \\
              =&4a^2J_2(2ay)-4a^2J_0(2ay)+2a^2\big[2J_{6}(2ay)+2J_{10}(2ay)-4J_{8}(2ay)\big]\cos\left(8\arg\delta\right). 
\end{align}

Similarly, we can find~$B_2$ and $C_2$ by substituting the relevant Jacobi-Anger expressions in Eqs.~(\ref{eq:b_2}), (\ref{eq:c_2}) and simplifying as shown for~$A_2$.
\section{Details of Taylor Series expansion for the superposition of $n$ compass states}
\label{section:taylorSeriesdetails}
For the superposition of an arbitrary number~$n$ of compass states, the overlap~\eqref{eq:overlapn} vanishes when,
\begin{align}
\label{eq:appgeneraloverlap}
    & \sum_{m=0}^{n-1}\Big[\cos\big(2a|\delta|\cos\left(\text{arg}\delta+\nicefrac{m\pi}{2n}\right)\big) +\cos\big(2a|\delta|\sin\left(\text{arg}\delta+\nicefrac{m\pi}{2n}\right)\big)\Big]\equiv0.
\end{align}
To obtain approximate solutions, we assign the direction of phase-space displacement as~$\text{arg}\delta$ and assign the size~$a$ pertaining to both compass states in the superposition. We then calculate the Taylor series for Eq.~\eqref{eq:appgeneraloverlap} around some~$y\in\mathbb{R}$. We truncate the Taylor series to quadratic order. The Taylor series of the first term in Eq.~\eqref{eq:appgeneraloverlap} is
\begin{align}
\cos\big(2a|\delta|\cos\left(\arg\delta+\nicefrac{m\pi}{2n}\right)\big) = &\cos\big(2ay\cos\left(\arg\delta+\nicefrac{m\pi}{2n}\right)\big)-\sin\big(2ay\cos\left(\arg\delta+\nicefrac{m\pi}{2n}\right)\big)2a\cos\left(\arg\delta+\nicefrac{m\pi}{2n}\right)\left(|\delta|-y\right) \nonumber \\
&-\cos\big(2ay\cos\left(\arg\delta+\nicefrac{m\pi}{2n}\right)\big)2a^2\cos^2\left(\arg\delta+\nicefrac{m\pi}{2n}\right)\left(|\delta|-y\right)^2 
\end{align}
and the Taylor series of the second term in Eq.~\eqref{eq:appgeneraloverlap} is
\begin{align}
\cos\big(2a|\delta|\sin\left(\arg\delta+\nicefrac{m\pi}{2n}\right)\big) = &\cos\big(2ay\sin\left(\arg\delta+\nicefrac{m\pi}{2n}\right)\big)-\sin\big(2ay\sin\left(\arg\delta+\nicefrac{m\pi}{2n}\right)\big)2a\sin(\arg\delta+\nicefrac{m\pi}{2n})(|\delta|-y) \nonumber \\
&-\cos\big(2ay\sin\left(\arg\delta+\nicefrac{m\pi}{2n}\right)\big)2a^2\sin^2(\arg\delta+\nicefrac{m\pi}{2n})(|\delta|-y)^2.
\end{align}
 The Taylor series of Eq.~\eqref{eq:appgeneraloverlap} is the sum of both these series.
\section{Form of quadratic coefficients for superposition of $n$ compass states}
\label{section:simplify_JA}
The coefficients of the quadratic equation~$(A_n(|\delta|)^2+B_n(|\delta|)+C_n)$ for a fixed~$\arg\delta$ and~$a$ for the region~$||\delta|-y|\ll1$ are
\begin{align}
\label{eq:general_a}
     A_n(y;a,\arg\delta) =&\sum_{m=0}^{n-1}\Big[-2a^2\sin^2\left(\arg\delta+\nicefrac{m\pi}{2n}\right) \cos\big(2ay\sin\left(\arg\delta+\nicefrac{m\pi}{2n}\right)\big) \nonumber \\
     &-2a^2\cos^2\left(\arg\delta+\nicefrac{m\pi}{2n}\right) \cos\big(2ay\cos\left(\arg\delta+\nicefrac{m\pi}{2n}\right)\big)\Big]
\end{align}
for the quadratic term,
\begin{align}
\label{eq:general_b}
     B_n(y;a,\arg\delta) = &\sum_{m=0}^{n-1} \Big[4a^2y\sin^2\left(\arg\delta+\nicefrac{m\pi}{2n}\right)\cos\big(2ay\sin\left(\arg\delta+\nicefrac{m\pi}{2n}\right)\big) \nonumber \\
     &-2a\sin\left(\arg\delta+\nicefrac{m\pi}{2n}\right)\sin\big(2ay\sin\left(\arg\delta+\nicefrac{m\pi}{2n}\right)\big)  \nonumber \\&
    + 4a^2y\cos^2\left(\arg\delta+\nicefrac{m\pi}{2n}\right)\cos
    \big(2ay\cos\left(\arg\delta+\nicefrac{m\pi}{2n}\right)\big) \nonumber \\
    &-2a\cos\left(\arg\delta+\nicefrac{m\pi}{2n}\right)\sin\big(2ay\cos\left(\arg\delta+\nicefrac{m\pi}{2n}\right)\big)\Big]
\end{align}
for the linear term, and 
\begin{align}
\label{eq:general_c}
 C_n(y;a,\arg\delta) =  &\sum_{m=0}^{n-1}
         \Big[2ay\sin\left(\arg\delta+\nicefrac{m\pi}{2n}\right)\sin\big(2ay\sin\left(\arg\delta+\nicefrac{m\pi}{2n}\right)\big) \nonumber \\
         &-2a^2y^2\sin^2\left(\arg\delta+\nicefrac{m\pi}{2n}\right) \cos\big(2ay\sin\left(\arg\delta+\nicefrac{m\pi}{2n}\right)\big)  +\cos\big(2ay\sin\left(\arg\delta+\nicefrac{m\pi}{2n}\right)\big) \nonumber \\ &+ \cos\big(2ay\cos\left(\arg\delta+\nicefrac{m\pi}{2n}\right)\big) 
         + 2ay\cos\left(\arg\delta+\nicefrac{m\pi}{2n}\right)\sin\big(2ay\cos\left(\arg\delta+\nicefrac{m\pi}{2n}\right)\big)\nonumber \\ &-2a^2y^2\cos^2\left(\arg\delta+\nicefrac{m\pi}{2n}\right) \cos\big(2ay\cos\left(\arg\delta+\nicefrac{m\pi}{2n}\right)\big)\Big].
\end{align}
for the constant.
We further simplify Eq.~(\ref{eq:general_a}) to
\begin{align}
\label{eq:intermediateAn}
        A_n(y;a,\arg\delta) = &\sum_{m=0}^{n-1} \Big[-2a^2\sin^2(\arg\delta+\nicefrac{m\pi}{2n}) \cos\big(2ay\sin\left(\arg\delta+\nicefrac{m\pi}{2n}\right)\big) \nonumber \\
     &-2a^2\cos^2(\arg\delta+\nicefrac{m\pi}{2n}) \cos\big(2ay\cos\left(\arg\delta+\nicefrac{m\pi}{2n}\right)\big)\Big] \nonumber \\
              =&\sum_{m=0}^{n-1}\Big[-a^2\big(1-\cos2(\arg\delta+\nicefrac{m\pi}{2n})\big)\cos\big(2ay\sin\left(\arg\delta+\nicefrac{m\pi}{2n}\right)\big) \nonumber \\
              &-a^2\big(1+\cos2\left(\arg\delta+\nicefrac{m\pi}{2n}\right)\big)\cos\big(2ay\cos\left(\arg\delta+\nicefrac{m\pi}{2n}\right)\big)\Big] \nonumber \\
              =&\sum_{m=0}^{n-1}-\Big[a^2\Big(\cos\big(2ay\sin\left(\arg\delta+\nicefrac{m\pi}{2n}\right)\big)+\cos\big(2ay\cos\left(\arg\delta+\nicefrac{m\pi}{2n}\right)\big)\Big) \nonumber \\
              &+a^2\cos2(\arg\delta+\nicefrac{m\pi}{2n})\Big(\cos\big(2ay\sin\left(\arg\delta+\nicefrac{m\pi}{2n}\right)\big)-\cos\big(2ay\cos\left(\arg\delta+\nicefrac{m\pi}{2n}\right)\big)\Big)\Big]. 
\end{align}
\par
Further simplifying with the help of Eq.~\eqref{eq:JacobiAnger} we obtain
\begin{align}
    \cos\big(2ay\sin\left(\arg\delta+\nicefrac{m\pi}{2n}\right)\big)+\cos\big(2ay\cos\left(\arg\delta+\nicefrac{m\pi}{2n}\right)\big) =& 2J_0(2ay)+4\sum_{s=1}^{\infty}J_{4s}\left(2ay\right)\cos\big(4s\left(\arg\delta+\nicefrac{m\pi}{2n}\right)\big) 
\end{align}
and
\begin{align}
    \cos\big(2ay\sin\left(\arg\delta+\nicefrac{m\pi}{2n}\right)\big)-\cos\big(2ay\cos\left(\arg\delta+\nicefrac{m\pi}{2n}\right)\big) =& 4\sum_{s=1}^{\infty}J_{4s-2}(2ay)\cos\big(\left(4s-2\right)\left(\arg\delta+\nicefrac{m\pi}{2n}\right)\big). 
\end{align}
\par
Inserting these back in Eq.~\eqref{eq:intermediateAn} yields
\begin{align}
    A_n(y;a,\arg\delta=&2a^2nJ_2(2ay)-2a^2nJ_0(2ay)+\sum_{m=0}^{n-1}\sum_{s=1}^{\infty}\Big[\big(2J_{4s-2}(2ay)+2J_{4s+2}(2ay)-4J_{4s}(2ay)\big)a^2\cos\big(4s\left(\arg\delta+\nicefrac{2\pi m s}{n}\right)\big)\Big] \nonumber \\
              =&2a^2nJ_2(2ay)-2a^2nJ_0(2ay)+na^2\big(2J_{4n-2}(2ay)+2J_{4n+2}(2ay)-4J_{4n}(2ay)\big)\cos\big(4n\left(\arg\delta\right)\big). 
\end{align}

Similarly, we can find~$B_n$ and $C_c$ by substituting the relevant Jacobi-Anger expressions in Eqs.~(\ref{eq:general_b}), 
 (\ref{eq:general_c}) and simplifying as shown for~$A_n$. 
\end{widetext}

\bibliography{trial.bib}
\end{document}